# Experiment indicates that Realism, not Locality, is false in Quantum Mechanics.

Mónica Agüero, Juliana Bourdieu, Alejandro Hnilo, Marcelo Kovalsky and Myriam Nonaka.
*CEILAP, Centro de Investigaciones en Láseres y Aplicaciones, (MINDEF-CONICET);*
*J.B. de La Salle 4397, (1603) Villa Martelli, Argentina.*
*alex.hnilo@gmail.com*

*Abstract*

The interpretation of the meaning of Quantum Mechanics has faced controversy since its inception. Bell's inequalities are a touchstone in this controversy. Their observed violation demonstrates that at least one of the hypotheses involved in their derivation and test is false in Nature. In principle, one has to choose between accepting that Locality is false, what implies a possible contradiction with the Theory of Relativity, or accepting that Realism is false, what means to give up the existence of a physical world independent of the observer. The right answer has consequences both foundational and practical, and theoretical discussions have searched it for decades. We report the results of a Bell's experiment designed and performed to add observational information to the discussion. Three proposals to reveal the false hypothesis are carried out, namely: search of attractors in time series of observations, variation of randomness of binary series of outcomes between space-like and not-space-like separated conditions of observation, and test of a bound of Kolmogorov's complexity. The results are consistent with the absence of the specific form of Realism usually involved in the derivation of Bell's inequalities, while remaining compatible with Locality within the sensitivity of the tests. Independently of the foundational problem and of any interpretation, some bare observations have immediate practical impact on the best use of device-independent quantum Random Number Generators and Quantum Key Distribution.

## Introduction.

Quantum Mechanics (QM) currently celebrates its centenary. In spite of its absolute success in predicting observable results, some issues regarding its meaning remain unclear. During its early years, the nature of quantum predictions was intensely debated between A.Einstein and N.Bohr [1]. Einstein objected quantum intrinsic indeterminacy and claimed QM to be an incomplete description of physical reality. The famous Einstein-Podolsky-Rosen (EPR) paradox was a climactic point in this debate, pointing out a contradiction with some apparently indisputable features of the physical world. Three decades later, refinement of the EPR argument led to experimentally testable predictions, the Bell's inequalities (BI). These inequalities are derived from the intuitive hypotheses of "Locality" and "Realism" [2]. Recent observed violation of BI in conditions practically free of relevant experimental imperfections ("loophole-free" [3-8]) implies that at least one of these intuitive hypotheses is false [9]. The question is: which one? The answer has not only foundational, but also practical consequences.

In short, "Locality" means that there are no influences propagating faster than the speed of light. In consequence, the outcome of a measurement cannot depend on whatever is outside its past light cone. Supposing that Locality is the false hypothesis opens the door to conflict with the Theory of Relativity. The conflict is usually circumvented by requiring that these influences cannot transmit information. Yet, the tension remains. That's why A.Shimony famously stated that QM and Relativity are in "peaceful coexistence", a term which had been coined to describe the fragile relationship between the East and West blocks during the Cold War.

Also in few words, "Realism" is the idea that the physical world exists independently of being observed ("Is the Moon there when nobody looks?" [10]). It is probably the most basic scientific postulate imaginable.

Having to choose between falsity of Locality and falsity of Realism is therefore an uncomfortable situation. It is one of the oddest features of QM. On the one hand, there is a widespread belief on "quantum non-Locality". Wording often assumes violation of BI and non-Locality to be equivalent [11,12]. On the other hand, many distinguished researchers have insisted, with different arguments, that the belief in quantum non-Locality is misled [13-18].

The precise definitions of Locality and Realism is a complex issue [10,19,20], which discussion is beyond the scope of this contribution. Just to avoid confusion, *in this paper* they are the ones involved in the usual derivation of the BI [2], that is:

Realism: the probability of an observable's outcome is given by a well-behaved integral of classical probabilities $P(\lambda)$ and distributions $\rho(\lambda)$ over a space of hidden variables $\lambda$. F.ex., the probability of observing the outcome "1" in station A when the setting is $\alpha$ (see Figure 1) is:

$$P_A^1(\alpha)= \int d\lambda.\rho(\lambda).P_A^1(\alpha,\lambda) \quad (1)$$

Be aware this definition does not include all conceivable forms of Realism, f.ex., the ones that do not involve the use of classical probabilities, and/or Bohmian Mechanics.

Locality: effects propagating faster than light do not exist. In consequence, probabilities of events occurring in *space-like separated* conditions must be statistically independent. F.ex, the probability of observing a coincident detection A=1 and B=0 when the settings are {α,β} in the setup in Fig.1 is:

$$P_{AB}^{10}(\alpha,\beta,\lambda)= P_A^{1}(\alpha,\lambda).P_B^{0}(\beta,\lambda) \qquad (2)$$

where ρ(λ) is assumed independent of {α,β} ("non-contextuality") and {α,β} independent of λ ("measurement independence"). In order to enforce validity of Eq.2 in practice, the "loophole free" experiments were performed at large distances between stations and fast and arbitrary variation of the settings {α,β}, what meant a formidable technical achievement.

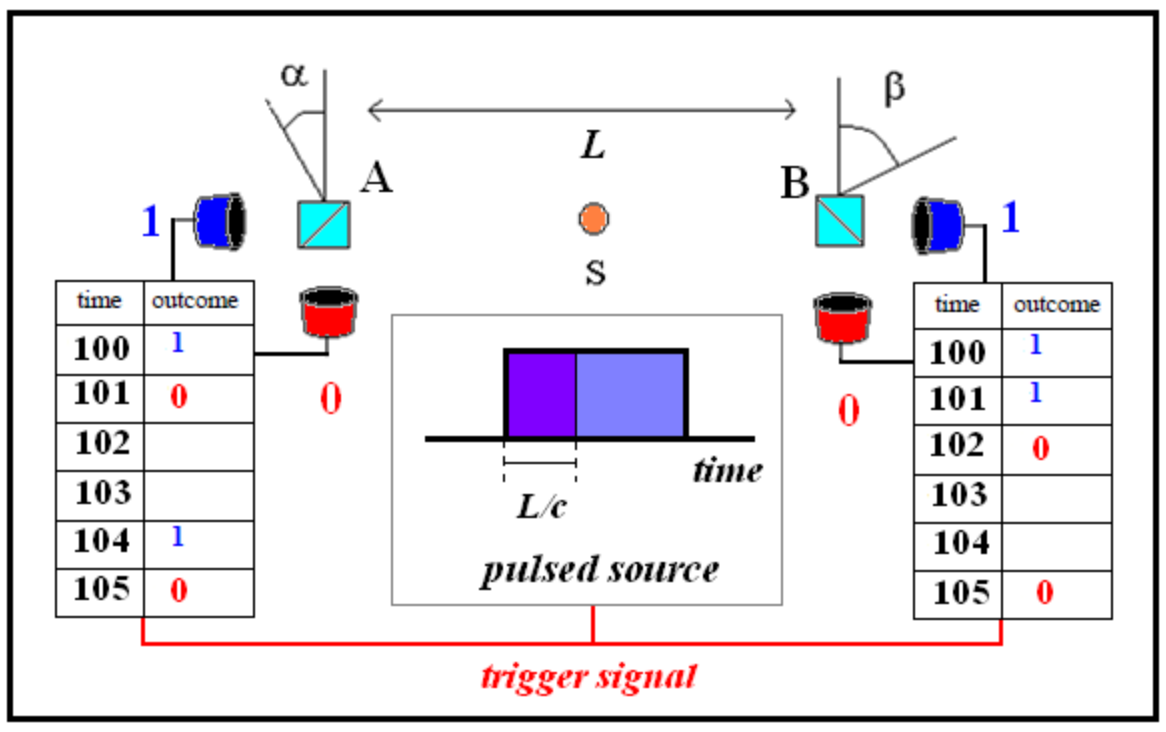

Figure 1: Sketch of a pulsed Bell's setup generating time and binary series. Source S emits states entangled in polarization to stations A and B, separated by distance *L*. In the tables, "1" (0) means that detector "1" (0) fired during that time slot, a space in blank means that no detector fired. A coincidence (1,1) occurs at slot #100, a (0,1) at #101, a (0,0) at #105. Only coincidences are taken into account. The setup generates two types of series: outcomes (binary), and time values. The binary series at A is (1,0,0) at B is (1,1,0); the time values' series is (100,101,105). Coincidences occurring in time slots during the pulses' earlier (darker) area are made of space-like separated events. They form series which, depending on which hypothesis is false, are expected to have average randomness or complexity values different from those recorded during the later (lighter, not space-like separated) area.

Leaving aside the foundational problem, whether Locality or Realism is false has different practical consequences. The idea of Quantum Certified Randomness (QCR) and the "no-signaling" theorems, relevant in the field of Quantum Information, are derived from assuming non-Locality as an axiom [21].

There is a complication: *experiments* devoted to test the BI need one additional hypothesis. The problem is that data for different setting angles (say, α and α') are necessarily recorded during different periods of time. Consider, f.ex., the derivation of the Clauser-Horne (CH) inequality [2]. Using Locality and Realism as defined in Eqs.1-2, plus a simple arithmetical property, one gets:

$$P_{AB}(\alpha,\beta,\lambda)+ P_{AB}(\alpha',\beta,\lambda) + P_{AB}(\alpha',\beta',\lambda) - P_{AB}(\alpha,\beta',\lambda) - P_B(\beta,\lambda) - P_A(\alpha',\lambda) \leq 0 \qquad (3)$$

where all probabilities correspond to detection in the detectors "1" in Fig.1. Applying ∫dλ.ρ(λ), the usual CH inequality is obtained. However, real measurements are not integrals on λ, but integrals on *time*. Each term in Eq.3 requires measuring during a different time interval, for it is not possible to set α and α' (or β and β') simultaneously. This impossibility breaks the logical chain of derivation. The physically sounding solution to retrieve the usual BI is to hypothesize that time averages, which are the ones actually recorded during experimental runs, are equal to the averages over hidden variables that are used in the derivation of the BI, f.ex:

$$\int dt.\rho(\lambda[t]).P_{AB}^{10}(\alpha,\beta,\lambda[t]) = \int d\lambda.\rho(\lambda).P_{AB}^{10}(\alpha,\beta,\lambda) \qquad (4)$$

This means equality between time and ensemble averages; that's why this additional hypothesis was originally named "Ergodicity" by V.Buonomano [22]. Its necessity for the experiments aimed to test the BI was often forgotten, and rediscovered under different names [23-27]. It is reviewed in the Supplementary Material Section 2.2.

Locality, Realism and Ergodicity are therefore three hypotheses on an equal footing, necessary to derive and observe the BI. When an experiment violates the BI, at least one of these hypotheses must be false. Deciding which one is false is an issue of importance, both foundational and practical. Theoretical discussions have lasted for decades without reaching consensus. Adding experimental information may help to identify the false hypothesis.

In this paper, we report the realization of the first experiment aimed at shedding new light on the issue in the form of observational data. We follow three proposals put forward by different Authors. In the next Section 1, these proposals are reviewed. In Section 2, the experiment is described. Results are presented in Section 3. No evidence is found that Locality or Ergodicity are false. As it will be discussed, the conclusion consistent with all recorded evidence is that Realism (as defined in Eq.1) is false. But be aware that this does not mean (and we do not claim) that falsity of Realism is demonstrated. It is just *indicated* as the most likely one among several alternatives.

In order to avoid any possible confusion, we warn the Reader that the experiments reported here are *not* tests of the violation of BI. Violation of BI, non-signaling and parameter independence are assumed facts. Marginal probabilities are neither measured nor involved in the reasoning. Probability distributions are not considered here at all, except to calculate minimum entropy. Instead, measurements and reasoning involve low dimensional objects in phase space, series' randomness and complexity.

## 1. Description of the three proposals.

*1.1 First proposal: search of attractors.*

J.Hansson proposed to decide between Locality and Realism by searching the presence of compact objects of finite dimensionality, or *attractors*, in the phase space of *series of time values* of quantum observations [28]. In short: if an attractor is found, then quantum outcomes are determined by an underlying dynamic mechanism. In consequence, objective (classical) reality exists, then Realism is true, and Locality must be false. If no attractor is found instead, then the underlying dynamics does not exist, Realism is false and Locality must be true.

Hannson did not consider the possibility of non-Ergodicity. The existence of an attractor is often considered as evidence of non-Ergodicity, for it suggests a metrically decomposable phase space (see the Supplementary Material Section 2.2). Therefore, observation of an attractor indicates that Realism is valid, but the consequence can be non-Locality or non-Ergodicity (or both). The third proposal (see Section 1.3 below) is a direct test of Ergodicity, and is hence complementary to Hannson's proposal.

Takens' embedding theorem states that the structure of the attractor can be revealed, no matter what is the original phase space, in a vector space of dimension dE reconstructed from a *time* series of scalar measurements $\{x(1), x(2), \ldots, x(N)\}$. This is done by building sets of vectors of the form [29]:

$$\{x(n), x(n+m), x(n+2m), \ldots, x(n+(\mathrm{dE}-1)m)\} \quad (5)$$

The value of the delay $m$ is usually chosen equal to the first minimum in the average mutual information calculated along the series.

The number of dimensions dE necessary to unfold the attractor is found by eliminating the apparent overlaps existing in a reconstruction of insufficient dimension. F.ex.: let suppose that the attractor is a cube. In a one-dimension reconstruction, a segment of the line is obtained. Many points of the cube, that are actually distant, appear close to each other within that segment of the line. In a two-dimension reconstruction instead, many of the former near neighboring points are revealed to be actually distant. They were *false nearest neighbors* (*fnn*) in the one-dimension reconstruction. When going to a three-dimension reconstruction, even more points are revealed to have been *fnn*, for the same reason. Yet, going beyond three dimensions does not reduce the number of *fnn* further, for the cube *is* an object in three dimensions. The unfolding is complete.

The practical procedure to reveal the presence of an attractor is like applying a sequence of filters. The first filter is the smooth decay of the fraction of *fnn* to zero as dE is increased in eq.5. The second filter involves shuffling ("surrogate") the series. To pass this filter, the *fnn* of the surrogate series should *not* decay to zero. The final filter is to reconstruct the attractor and to predict elements in the series. If the series does not pass one of the filters, then it is concluded there is no attractor present in that series.

Attractors were searched in this way in 365 time series of different lengths recorded in the famous Bell's experiment performed in Innsbruck [30]. An attractor was unquestionably found in the longest series in real time [31]. According to Hannson's proposal, this result is evidence of non-Locality. Yet, in the opinion of the first Author of that experiment (G.Weihs) that result was probably caused by a drift between the clocks in the remote stations. By then the setup had been dismantled, so it was impossible to know the cause for sure. The issue remained undecided. It is fair to recall that that experiment had been designed and performed for a different purpose (i.e., closing the predictability or measurement independence loophole, what it did).

Practical hints for attractor's reconstruction from experimental data are explained in the Supplementary Material Section 4.

In the experiment reported in this paper, stations are placed at $L$=24m so that early detections occur as space-like separated events. This enforces the violation of at least one of the three hypotheses. Data with stations close ($L$=1.5m) are also recorded, to act as reference. Due to instruments' response timescale, detections recorded when $L$=1.5m are not space-like separated, and BI can be violated even if the three hypotheses are true. Therefore, if Realism is true, attractors should be found easier, or more often, in time series recorded when $L$=24m than when $L$=1.5m.

### *1.2 Second proposal: time variation of randomness.*

This proposal is based on the relationship between falsity of each one of the three hypotheses and randomness of the binary *series of outcomes* produced in a setup as in Fig.1. This relationship has been discussed at length in [32], and is reviewed here in the Supplementary Material Section 2. In what follows we present a short version of the reasoning.

If BI are violated because non-Local effects exist, binary series recorded in space-like separated conditions *must* be random. Otherwise, the non-Local effects could be used for sending signals at a speed faster than light, in open contradiction with the Theory of Relativity [33]. This is the basis of QCR. Note that, according to this reasoning, series made of not-space-like separated detections are not "random certified" even if they violate the BI, because some interaction propagating at subluminal speed may create (spuriously) correlations able to violate the BI, without necessarily generating random series.

Instead, if BI are violated because the hypothetical classical dynamics that generates the series (always recorded space-like separated) is non-ergodic (i.e., if Eq.4 is false), then the binary series in Fig.1 *must* be *not*-random. This issue is discussed in detail in the Supplementary Material Section 2.2. In few words, it can be visualized as follows: a classical system that evolves ergodically explores its (bounded) phase space evenly, that's why time and ensemble averages are equal [34]. If the evolution is non-ergodic instead, then the system spends more time in some regions of its phase space than in others of the same measure (for, time and ensemble averages are assumed not-equal). Hence, at a given time, the system is more probably found in some regions than in others. Its future state can be partially predicted. Therefore, an evolution that is non-ergodic is (at least partially) predictable and hence not-random, for all definitions of "random".

Nevertheless, we do not claim the relationship Random $\Rightarrow$ Ergodic to be general. It applies only to the case of interest here, that is, a classical dynamical system which generates binary series as it enters different regions of its bounded phase space. F.ex., the unbounded random walk is random by definition, and evidently non-Ergodic.

Finally, if the BI are violated because Realism is false, then there is no reason to say the binary series recorded in space-like separated conditions must be random, or not. In any case, not more (or less) random than series recorded in space-like *not* separated conditions. Note that Born's rule allows calculating the probability of an outcome, but is silent about the features of the series underlying the measurement of that probability.

According to the previous reasoning, it would suffice measuring the level of randomness of binary series of outcomes recorded in space-like separated conditions to reveal the false hypothesis. Unfortunately, "randometers" do not exist. There is not even a universally accepted *definition* of randomness. Martin-Löf's theorem ensures a universal test exists that determines if a given series is random, at least in the typical and algorithmic senses [35], but its expression is unknown. The problem of the meaning of "random" is complex; we discuss it in Supplementary Material Section 3 to not to interrupt the exposition of this proposal.

It is impossible measuring absolute randomness, but detecting *relative variations* may be at hand. For, it is reasonable assuming variations of standard evaluators of randomness (say, entropy) to be *at least coarsely correlated* with variations of "actual" randomness (say, as it would be given by the unknown Martin-Löf test). In other words, we assume the behavior of the evaluators (averaged over a significant number of series) to coarsely follow the behavior of "actual" randomness.

Let suppose then a pulsed Bell's setup as in Fig.1, and that violation of BI is constant during the pulses (as predicted by QM and also observed [36]). Detections during the first part of the pulses (i.e., when time $t<L/c$, being $t$ measured from the start of each pulse) occur as space-like separated events. If BI are violated during this period, this is possible only because Locality, or Realism, or Ergodicity is false. Series recorded during this first period must then have the corresponding level of randomness (respectively: high, indefinite, low). For $t>L/c$ instead, detections are no longer space-like separated and BI can be violated even if the three hypotheses are true. Series recorded during this second period can have any level of randomness. Therefore, changes of the randomness' evaluator (averaged over a large number of series) between series recorded at $t<L/c$ and $t>L/c$ indicates the false hypothesis. If the evaluator's average value decreases, it indicates that Locality is false (because series recorded during the first period *must* be random, but not necessarily in the second period). If that value increases instead, it indicates that Ergodicity is false (because series recorded during the first period *must* be not-random, but not necessarily in the second period). Finally, if the value remains constant, it indicates that Realism is false (because validity of Realism does not affect series' randomness). Note that the precise amount of the averaged evaluator's variation is irrelevant, all that matters is if it increases, decreases or remains constant.

It may be argued that a constant average value can be caused by non-Locality and non-Ergodicity exactly balancing each other. Yet, this alternative should be supported by the other proposals' results. In the same way, a (f.ex.) decreasing value (evidence of non-Locality) should be confirmed by the detection of attractors in the first proposal.

*1.3 Third proposal: lower bound of complexity.*

Let assume that the binary *series of outcomes* in Fig.1 are produced by an underlying classical dynamical mechanism, or *generator*. W.Zurek has demonstrated that, if the generator is ergodic, then the series' complexity $K$ is bound from below by the series' entropy $H$, i.e., $K \geq H$ [37].

Kolmogorov's complexity $K$ is defined as the length of the shortest program that reproduces the series. A series is defined as "algorithmically random" when $K$ is equal to the series' length. This is often believed to be the strongest form of randomness. Yet, $K$ cannot be computed. For, given a series and a program that reproduces the series, one can never be sure that there is no shorter program that does the same. What can be computed is an estimated complexity $Kc$ from the asymptotic compressibility of the series by using, f.ex., the algorithm devised by Lempel and Ziv [38]. As $K$ cannot be computed, Zurek's bound $K \geq H$ is most interesting. Taking into account that $Kc \geq K$ and $H \geq Hm$ (minimum entropy, see later), then:

$$Hm > Kc \Rightarrow \textit{Ergodicity is false} \qquad (6)$$

Both $Kc$ and $Hm$ can be calculated in experimentally recorded series of binary outcomes. Experimental confirmation of the bound $Kc \geq Hm$ in series produced in a Bell's setup, but not in space-like separated conditions, was reported in [39]. As before, violation of the bound is expected to be observed easier and/or more often in series recorded when $L$=24m than when $L$=1.5m. This proposal deals with Ergodicity only but, by combining its result with the results of the other two proposals, it helps to find which hypothesis is most likely to be false.

**2. Experimental setup.**

Figure 2 is the setup's diagram. Here we present the information needed to understand the results; technical details to allow reproducing the experiment are described in the Supplementary Material Section 1.

Because of technical limitations, the ideal experiment sketched in Fig.1 is unreachable nowadays. One problem is that, due to detectors' relatively low efficiency (even cryogenic nanowires), loophole-free violation of the BI is possible with photons only by using Eberhardt's states, which produce strongly non-uniform series [40]. Using extractors of randomness is out of the question, for they would destroy the series' features that may reveal the false hypothesis. Setups exploiting entanglement swapping between photons and matter do use Bell states, but they currently have detections' rate too low to be suitable.

Another problem is to achieve fast and unpredictable setting changes, which are necessary to close the predictability or measurement-independence loophole. In addition to the technical difficulty, there is the logical problem of performing unpredictable setting changes. An intrinsically non-testable hypothesis is always involved, for the changes must be unpredictable *by the source* of entangled states, but *not by the observer* [41]. Both problems can be circumvented by assuming that any conspiratorial correlation between the stations vanishes when the source of entangled states is turned off, what in our case occurs during the dead time between the pump pulses.

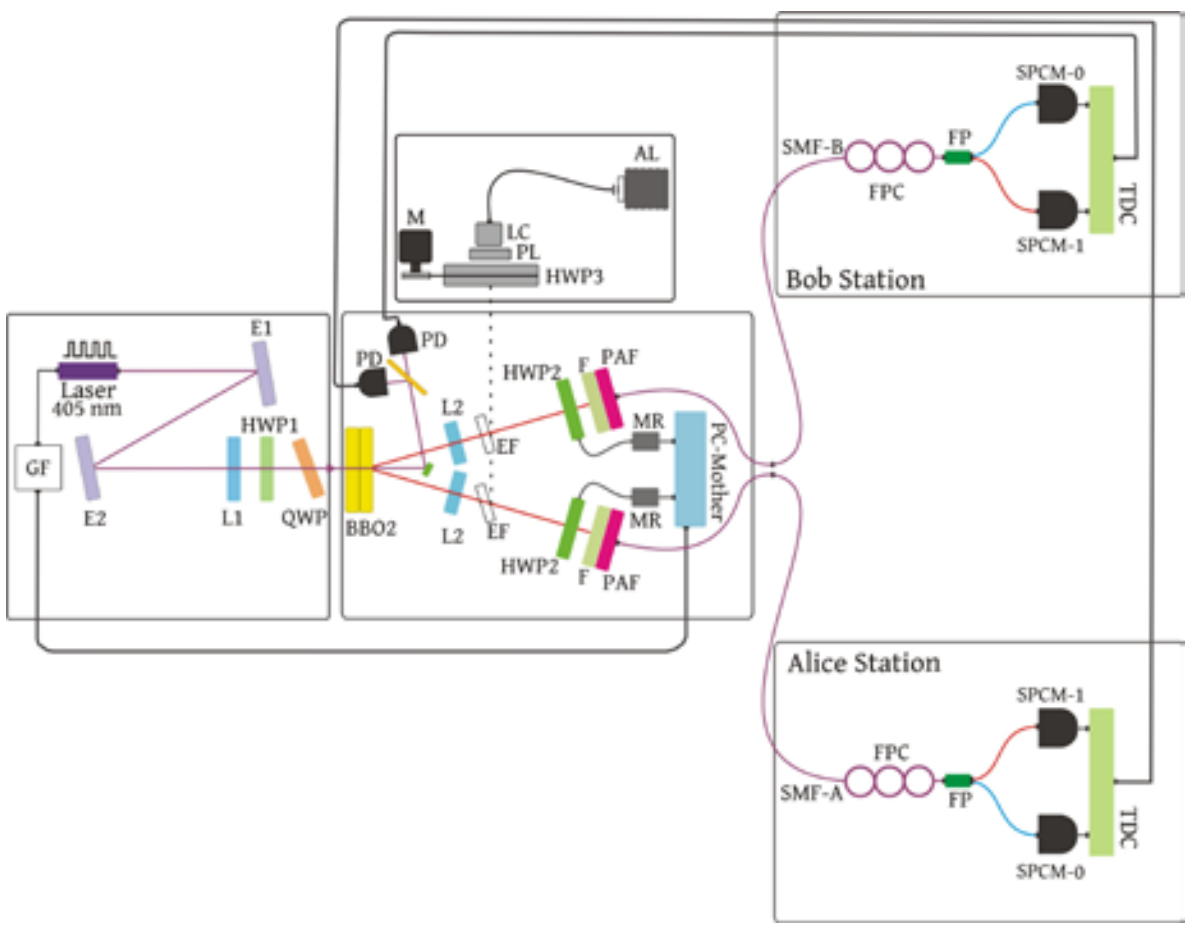

Figure 2: Diagram of the setup. It is a pulsed Bell's experiment, with time stamped record of data. GF: function generator that pulses the laser; L1,L2: f= 300 mm lenses; E1, E2: HR plane mirrors at 405 nm; HWP1 and QWP: half and quarter waveplates at 405 nm; BBO2: crossed BBO-I crystals (source of entangled states); TDC: time-to-digital converters, they record time values of photons' detections in channels #1 and #2, and samples from pump pulses in #3 as "trigger" signals; PD: fast photodiodes, they send samples of pulses' emissions to the TDCs via coaxial 50 Ω cables and provide "logical synchronization" free of clocks' drift [41]; HWP2, HWP3: half-waveplates at 810 nm; F: Interferential filters at 810 nm, Δλ=10 nm, 90% transmission; EF: auxiliary HR plane mirrors in flip-flop mountings to insert the beam of the auxiliary laser into the fibers; AL: auxiliary CW laser diode at 810 nm coupled to a multi-mode fiber, it facilitates birefringence compensation; LC: collimating optics; PL: linear polarizer; M: motor that rotates HWP3; MR: servo motor controllers of HWP2; PAF: fiberports f=7.5mm; SMF-A and B: single-mode fiber coils, 21m fixed length each; FPC: birefringence compensator ("bat-ears"); FP: fiber polarization analyzer; SPCM: photon counting module. The stations can be placed at different distances in straight line leaving everything else identical, which is a unique feature of this setup.

The experiment we report here entails hence two assumptions: *i)* Fair sampling [2]; i.e.: detectors do not conspire to detect (or not to detect) a photon according to the value of some hidden variable; *ii)* Any hypothetical interchange of conspiratorial information between the stations can occur only while the pump pulses are "on", and any resulting correlation vanishes after the end of each pulse. Noteworthy, independent observations support these assumptions, see Supplementary Material Section 1.3.

In our setup, a diode laser is pulsed at 500 kHz emitting square shaped pump pulses of duration 500ns. Photons entangled in polarization in the fully symmetrical Bell's state at 810nm propagate through single-mode optical fibers up to two stations (Alice and Bob) mounted on wheeled tables, which can be placed at adjustable distance *L* in straight line. This is a unique feature of our setup. In each station, a fiber polarizer separates the beams to silicon avalanche photodiodes.

Before starting a run, an auxiliary CW laser beam at 810 nm is inserted into the fibers through high reflectivity mirrors EF on flip-flop mountings (see Figure 2). Birefringence on the long fibers (SMF-A and B, 21m long each) is compensated by adjusting the bat-ears to maximize contrast between the two output gates of the fiber polarizers. For this action the SPCM are replaced by standard photodiodes. In order to facilitate maximizing the contrast, a half waveplate (HWP3) in a rotating mount (≈100 Hz) is inserted on the auxiliary laser beam before the EF mirrors. This produces a sinusoidal signal in the standard photodiodes after the polarizers, what facilitates contrast adjustment. Once birefringence distortions are satisfactorily compensated the flip-flops with the EF mirrors are moved "down" to allow the entangled photons propagate into the fibers, and the auxiliary laser is turned off.

Time to digital converters (TDCs) record detection time values of each photon. Samples of the pump pulses are sent as trigger signals to the stations, providing "logical synchronization" to the TDCs [42]. Besides, pulse frequency is modulated to define pulse numbering [43]. These innovations determine delay values and coincidences unambiguously, hence solving the mentioned problem of clocks' drift.

Time resolution of the TDCs is in the ps range, but in practice it is limited to 2ns because of detectors' jitter. Raw data have then the form of three files of time values: one for the pump pulse or trigger, and one for each detector, {1,0} in each station. Only one of ≈170 pump pulses produces a detected photon in the average. This is necessary in the pulsed regime to keep low the number of accidental coincidences [44]. These files are then processed, and time series of coincidences are obtained. Single detections are not taken into account.

Data can be then arranged in different ways, at will. Series of *time values* between successive coincidences are used for the first proposal, the search of attractors. Series of *outcomes* are binary, and are used to calculate randomness' evaluators *Kc* and *Hm* for the second and third proposals.

As said, we record series in two different cases: with the stations close (*L*=1.5m in straight line) and remote (*L*=24m). In each case 136 runs (each run means 30s of continuous observation in real time) amounting to ≈$4.1\times10^6$ coincidences, are collected. The *L*=1.5m case provides a reference, for at this short distance Locality, Realism and Ergodicity can all be true during the full pulses' duration and still the BI be violated (by some hypothetical mechanism propagating at sub-luminal speed). In the *L*=24m case instead, the series recorded during the early part of the pulses are obtained in space-like separated conditions, and have features (presence of an attractor, a distinctive level of randomness, violation of the $Kc \geq Hm$ bound) that should be absent in the remaining part of the pulses and in the whole *L*=1.5m case as well. Excepting the value of *L*, all experimental parameters including alignment, fibers' and cables' lengths, trigger levels in the TDCs, etc. are identical in both cases.

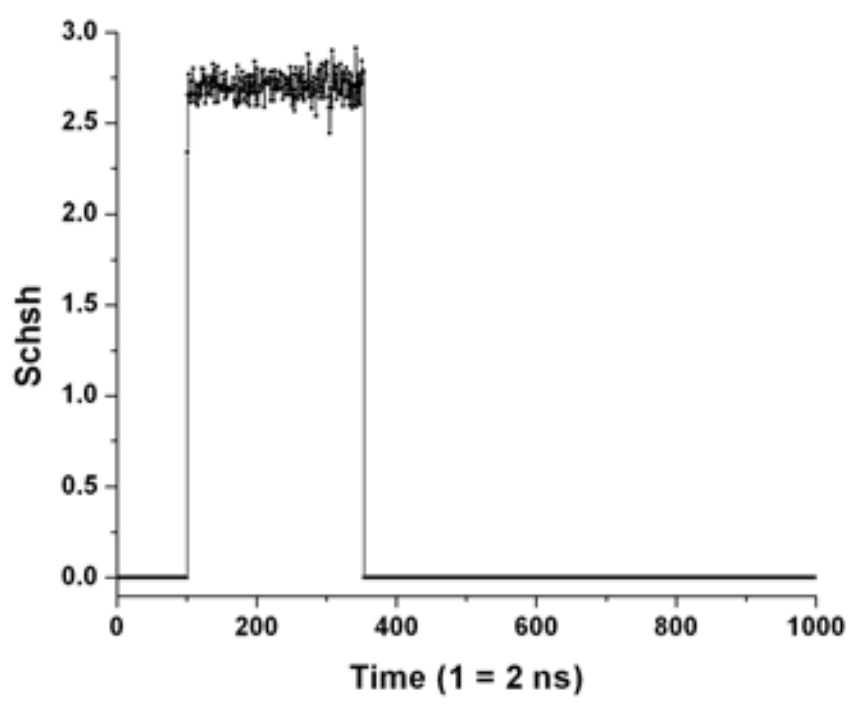

Figure 3: Evolution of $S_{CHSH}$ during the pulses, *L*=24m. Pulse duration 500ns, repetition rate 500 kHZ, coincidence window 2ns, full pulses' period is displayed, $\langle S_{CHSH}(t)\rangle = 2.73\pm0.07$.

The *same* recorded series allow reconstructing the evolution of the parameter $S_{CHSH}$ [2] during the pulses, see Figure 3 for *L*=24m. For *L*=1.5m it is practically the identical. $S_{CHSH}$ is constant and independent of the pulses' shape, as it has been already observed [36,45]. The relatively low contrast of fiber polarizers (<0.98 according to specs) limits the observable value of $S_{CHSH}$ to ≤ 2.77. We measure $\langle S_{CHSH}(t)\rangle = 2.73\pm0.07$.

In the next Section we discuss the results obtained according to each of the three proposals.

## 3. Results.

### *3.1 First proposal (attractors).*

As said, an attractor with dE=9 was found in one of the time series recorded in the Innsbruck experiment. That time series corresponded to 300s of continuous observation in real time, and accumulated 95797 coincidences. Here we record 64 time series, each with ≈$1.3\times10^5$ coincidences and 510s of observation in real time, which are hence comparable to the mentioned Innsbruck's time series.

The first filter to detect an attractor is that the *fnn* decay smoothly to zero as dE is increased. But, none of the 64 time series passes this first filter. They reach a minimum and then increase again. The curve *fnn* vs dE is plot in Figure 4 for a typical case (*N*=144182, *m*=1, first minimum of the mutual information). It means that no attractor exists. The analogous curve for the Innsbruck attractor (*m*=5, second minimum of the mutual information, see [31]) is also plot, to illustrate the difference.

As a further test, we compose series only made of coincidences occurring in space-like separated conditions. That is, series made of coincidences recorded only during the 80ns after the start of each pulse (*L*=24m). There are four of these series, each with ≈$1.5\times10^5$ coincidences. But, once again, no one passes the first filter. Besides, there is no noticeable difference between the curves of *fnn* vs dE drawn from the set of time series with *L*=24m and the ones recorded with *L*=1.5m.

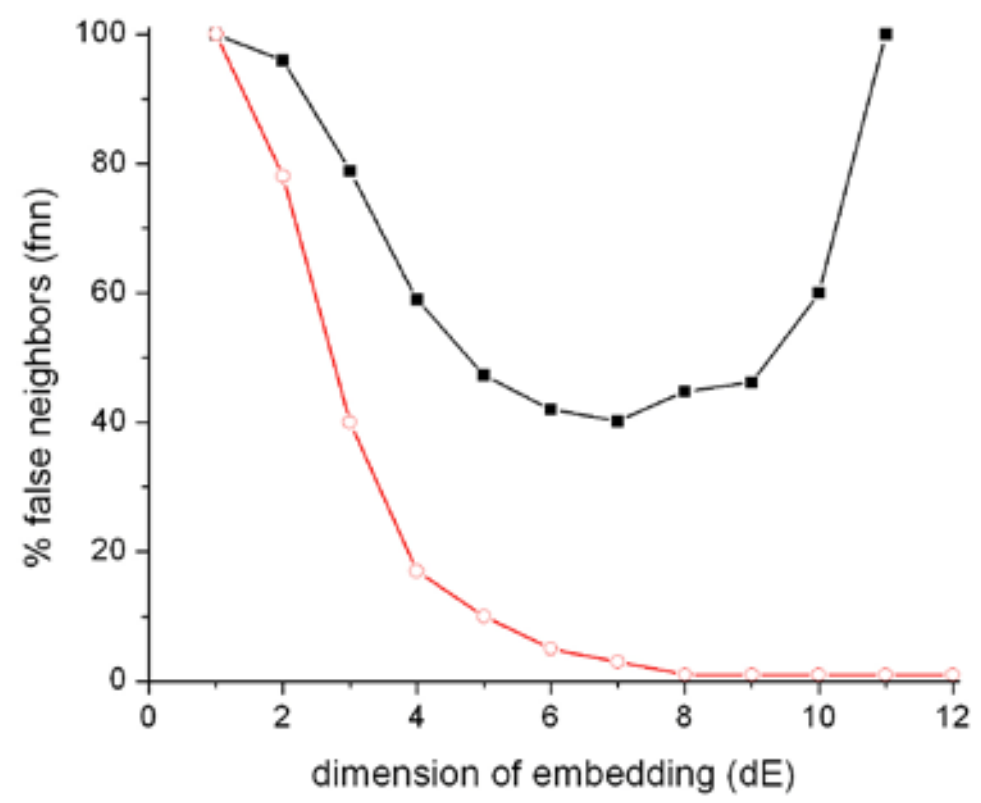

Figure 4: Curves *fnn* vs dE of series with and without an attractor. Open red circles: from data of the Innsbruck attractor (adapted from [31]). Full black squares: from one of the 68 comparable time series in this study.

In short, no decay of *fnn* to zero (hence no value of dE) is observed in any of the 68 recorded time series of length and duration comparable to the Innsbruck's one. No attractor is found. According to the reasoning in Hannson's proposal, Realism is false and Locality is true. Yet, an attractor may exist and pass unnoticed because of noise or insufficient data. The safe conclusion is that no evidence of non-Locality and non-Ergodicity is found (i.e., Locality and Ergodicity can be valid).

Thanks to the "logical synchronization" scheme, our setup is free of the mentioned clocks' drift, so our results suggest that the Innsbruck attractor was in fact caused by that drift, as G.Weihs had said.

### *3.2 Second proposal (randomness).*

We choose estimated and normalized Kolmogorov's complexity *Kc* and Minimum Entropy *Hm* as randomness' evaluators. We use the approach developed by Kaspar and Schuster [46] and implemented by Mihailovic [47] to calculate *Kc*. It is scaled from ≈0 for a regular series to ≈1 for an incompressible one. Minimum Entropy is usual in studies of quantum randomness because of a well-known relationship with $S_{CHSH}$ (see Supplementary Material Section 2.1). It is defined as:

$$Hm \equiv -\log_2 [max_r\, P(r)] \qquad (7)$$

where P($r$) is the probability of obtaining the outcome $r$ in the series. It also ranges between 0 and 1.

We stress that neither *Hm* nor *Kc* really measure randomness but, as said, it is reasonable to assume their averages over a statistically significant number of series to follow, at least coarsely, any relative variation of "actual" randomness that may occur.

We divide the pulses' duration into 5 "slots" of 100ns each. We compose series of binary outcomes from the coincidences detected within each slot; 136 series are generated in this way for each slot, each station and each value of *L*, so we get 2720 binary series in total. Each one is ≈6 Kbit long.

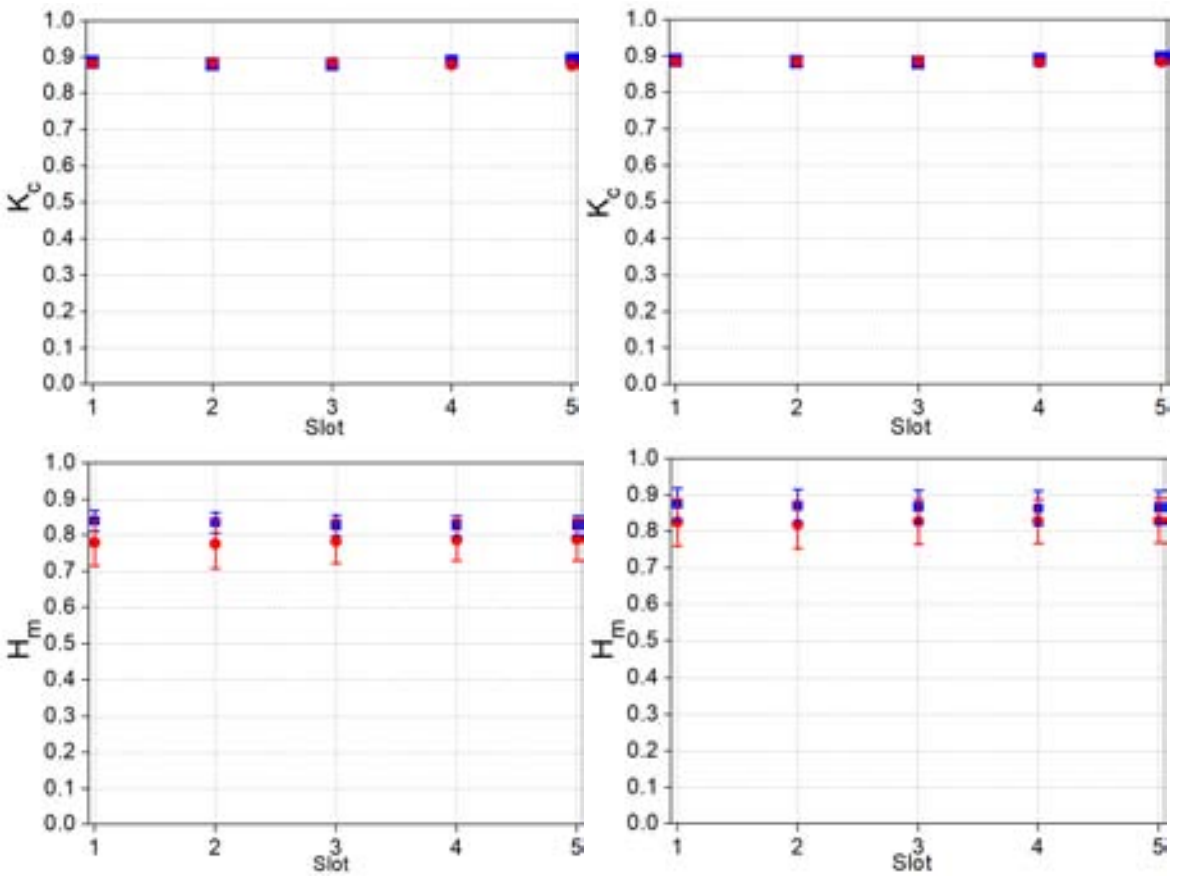

Figure 5: Averaged estimated complexity *Kc* (upper row) and Minimum Entropy *Hm* (lower row) of binary series recorded at Alice (left column) and Bob (right column) stations; each slot has 100ns duration. Red circles: *L*=24m, Blue squares: *L*=1.5m. Statistical dispersion over 136 series is indicated in each dot.

In Figure 5, evolution of *Kc* and *Hm* during the pulses are displayed, for both *L*=24m and *L*=1.5m cases and for both stations. Each dot corresponds to the average value of these evaluators of randomness over the set of 136 series; error bars indicate the dispersion in this set.

Series recorded in the first slot when *L*=24m (*L/c*≈80ns) are almost fully made up of space-like separated events. If the BI are violated because Locality or Ergodicity are false, then the average of the randomness' evaluators in these slots should be different from the averages in the subsequent 4 slots, and from the averages in all the slots when *L*=1.5m. But, no such differences are observed.

As a further test, we increase the number of slots to 10 and repeat the calculations. Now all series in the first slot are fully made up of space-like separated events. Now there are 5440 series, each ≈3 kbit long. Be aware that, because of the way they are composed, these series are completely different from the ones obtained with 5 slots (i.e., they are not merely the first of second half of the series obtained with 5 slots). The results are shown in Figure 6.

In both Figs. 5 and 6, it is visually evident there is no variation of *Hm* or *Kc* between the first slots and the subsequent ones. No different variations between the *L*=24 m and *L*=1.5 m cases are observed either. We analyze this issue in detail in the Supplementary Material Section 5 by applying Student's t-test with $\alpha$=0.95 statistical significance, among other criteria. The results of the detailed analysis concur with the visual evidence.

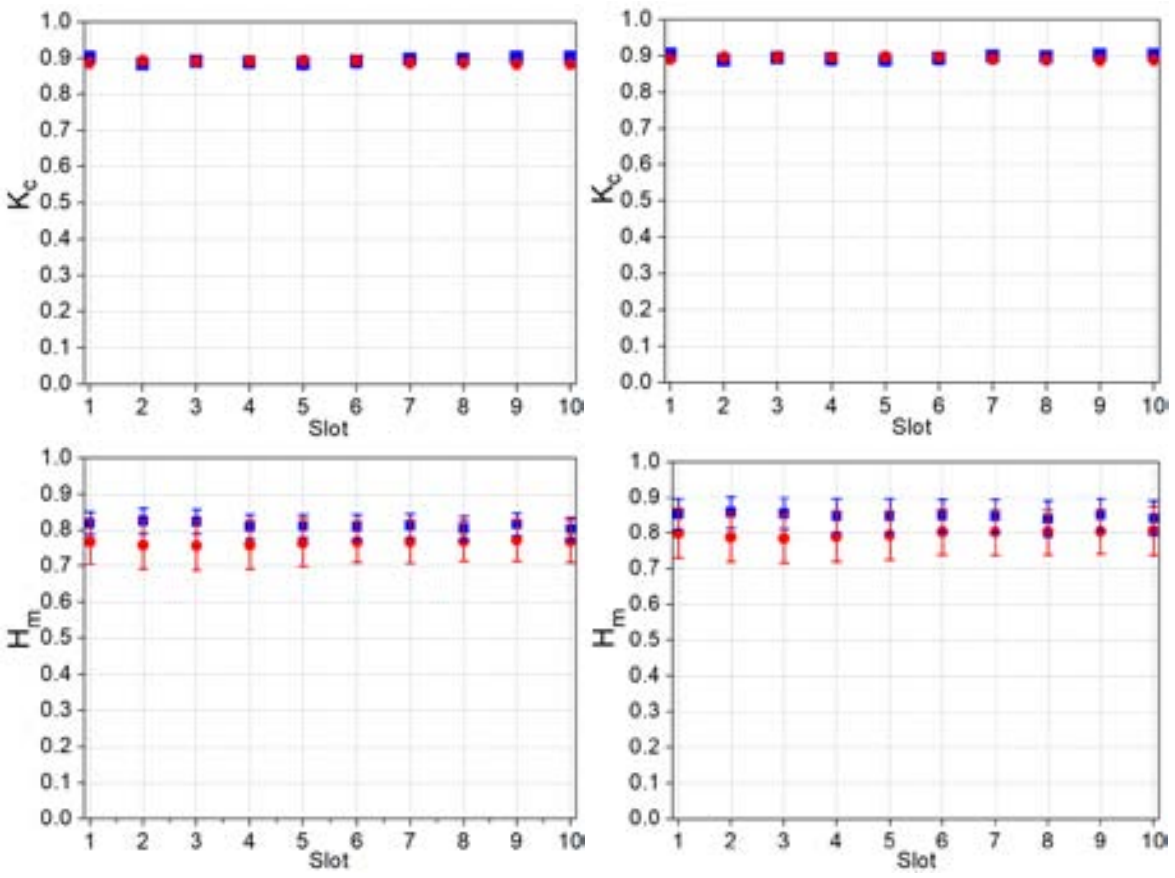

Figure 6: The same as Fig.5, but for slots 50 ns duration.

According to the reasoning in this second proposal, these results provide no evidence that Locality or Ergodicity are false. It is still possible *both* to be false simultaneously and to compensate each other exactly; but, in this case, an attractor should be observed in the first proposal, and non-Ergodicity should be revealed in the third proposal (unless, of course, that *both* experiments failed). Instead, figs.5 and 6 are consistent with what is *prima facie* expected if Realism is false.

### *3.3 Third proposal (complexity).*

According to this proposal, if Ergodicity is false, the bound *Kc*≥*Hm* should be violated in all the earlier slots when *L*=24m, and not violated (or, at least, less frequently violated) in the later ones, and in all slots when *L*=1.5m. The practical comparison between these two situations (i.e., how many times the bound is violated, if any) is then the key to reveal that Ergodicity is false.

Figure 7 plots the averages of *Hm vs Kc* over the same sets as in the second proposal. Red, full circles correspond to the early slots in the *L*=24m case. If Ergodicity is false they are expected to be, at least, closer to the boundary *Hm* = *Kc*, which is indicated with a dotted straight line. But, no values violate the bound regardless they correspond to series made of outcomes that occur space-like separated (red, full circles) or not (open circles). Besides, the red full circles as a whole are not closer than the others to the bound; the averages of both sets of circles are coincident. For the red full circles $\langle Kc\rangle$ = 0.889 ± 0.003, $\langle Hm\rangle$ = 0.807 ± 0.018, for the open circles $\langle Kc\rangle$ = 0.890 ± 0.001, $\langle Hm\rangle$ = 0.814 ± 0.004.

According to the framework of this third proposal, no evidence is found that Ergodicity is false. In consequence, the alternative that the results in Figs.5 and 6 are explained by simultaneous non-Locality and non-Ergodicity becomes less likely. Regarding the first proposal, this result is consistent with *not* detecting attractors.

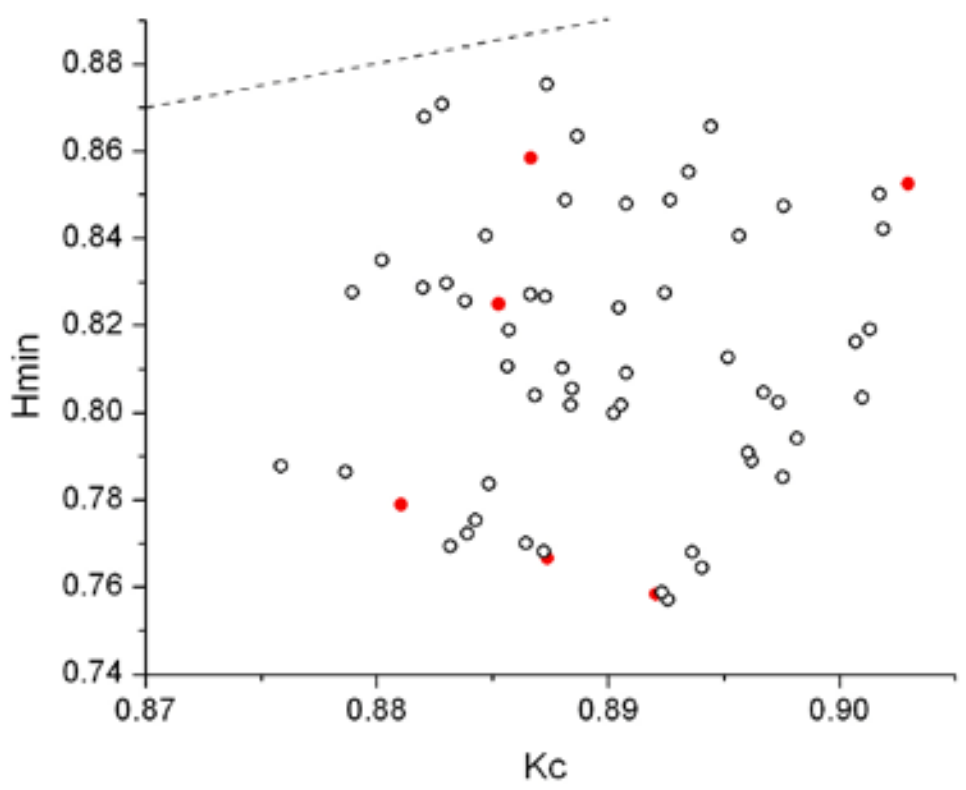

Figure 7: Values of *Hm* and *Kc*. Red (full) circles correspond to the early dots in the case *L*=24m, open circles correspond to all the other cases; the dotted line indicates the condition *Hm*=*Kc*. If the BI were violated because Ergodicity is false, the red circles should be above the dotted line or, at least, be closer to this line than the open circles.

**Summary and conclusions**.

Experimental evidence about which hypothesis is false (Locality, Realism or Ergodicity) when BI are violated is searched for the first time. Three different proposals by different Authors are carried out. No evidence is found that Locality or Ergodicity are false. Within the logical structure of the three proposals considered here, the most likely conclusion is that Realism (as defined in Eq.1) appears falsified.

Let see this conclusion in some detail: the results of the first proposal provide no evidence of Realism, and hence no evidence of non-Locality or non-Ergodicity. The ones of the second proposal are consistent with non-Realism and with *simultaneous* non-Locality and non-Ergodicity. The ones of the third proposal provide no evidence of non-Ergodicity. Therefore, non-Locality would be true only if *two* experiments (first and third) failed to reveal the evidence (attractor and bound violation) that was being searched. This is possible, of course (no experiment is perfect), but less likely than the alternative: non-Realism is consistent with the results of *all* three experiments even if none of them fails.

Because of the logic intrinsic in the proposals, the results cannot *demonstrate* that Locality and Ergodicity are true and that Realism is false. The absence of dynamical signatures does not exclude realist models that do not rely on underlying time dependent hidden-variable models, nor models in which such dynamics are not accessible with the present experimental resolution. Nevertheless, as said, the aim of this paper is just to add observational data to the controversy. These data might have been ambiguous or contradictory, but they point all in the same direction instead. To say the least, they tip the balance of the Realism vs. Locality (as defined in Eqs. 1 and 2) controversy in a clear way.

Recall that explicit rejection of non-Locality and a critic of reality independent of the observer was Bohr's early reaction to the EPR-paradox [48], setting the basis for the Copenhagen interpretation of QM.

Leaving aside the problem of revealing the false hypothesis, the mere recording of the time variation of averaged randomness' evaluators in Figs. 5 and 6 has immediate practical impact on the best use of device-independent QKD and Bell-based RNG. Independently of any interpretation, if *Hm* and *Kc* had been observed higher when t<*L/c* (no matter why) it would have been advisable using pump pulses shorter than *L/c* to get safer keys in QKD (in this case, *L* would have been the distance between emitter and receiver) and to obtain better sets of random series in RNG. If *Hm* and *Kc* had been higher when t>*L/c* instead, then it would have been advisable using the ending section of long pulses. But the evaluators' averages are observed to be constant in time. Therefore, pulse length and pulse section chosen are unimportant from this point of view. This information simplifies the implementation of those applications.

**Acknowledgments.**

This contribution received support of grant PIP 2022-0484CO from CONICET (Argentina).

**References.**

**SUPPLEMENTAL MATERIAL** for: "Experiment indicates that Realism, not Locality, is false in Quantum Mechanics"; M.Agüero, J.Bourdieu, A.Hnilo, M.Kovalsky and M.Nonaka.

**Index:**

## 1. Experimental setup.

*1.1 Description.*

Figure SM1 is a diagram of the experimental setup. Biphotons at 810 nm in the fully symmetrical entangled Bell state $|\varphi^+\rangle$ are produced in two crossed (1mm long each) BBO-I crystals, pumped by a pulsed diode laser at 405 nm. The average power of the pump laser at CW is 40 mW. Coherence length is measured 40 mm at 100 kHz pulse repetition rate and 10% duty cycle. This means $\Delta\omega_p \approx 5{\times}10^{10}$ s$^{-1}$, which is much smaller than the bandwidth of the spontaneous photon down conversion process in the crystals, and also than the filters' bandwidth $\Delta\omega_f \approx 3{\times}10^{13}$ s$^{-1}$. Coherence length is observed to increase with duty cycle (in CW operation it is 20 m according to specs). This laser is able to emit fairly square pulses of adjustable duration and repetition rate. Pulse shape deteriorates if the rate is higher than 1MHz or pulse duration is shorter than 200 ns. Duty cycles as low as 5% have been used with satisfactory results.

A sample of the pump beam is sent to a 50-50 beam splitter. The output beams illuminate two fast photodiodes, which send electrical signals to each station indicating the start of each pump pulse. These signals are sent through 50Ω coaxial cables 38 m length each, and are checked to have negligible distortion. Trigger signals are stored in the #3 input channels of the time-to-digital converters (TDCs, Id Quantique Id-900). These are the largest files, because most pulses are "empty": only ≈0.6% of the pulses produce detected photons. This is necessary to keep low the number of accidental coincidences in the pulsed regime [44]. In a typical run, tens of millions of trigger signals must be recorded correctly by both TDCs, what is challenging. In order to keep tracking of pulse numbering with independent clocks (which unavoidably drift away), the repetition rate is switched or modulated, in order to establish a "logical" synchronization between the clocks in the two TDCs [42,43]. The pulsed regime refreshes the synchronization between the distant clocks with each pulse; the frequency modulation allows reliable pulse numbering and immediately determines the correct delays between the lists of photons' detections, with no need of convoluting the time lists and counting coincidences for each possible value of delay. This is a significant practical advantage, and provides unambiguous and more reliable results.

The entangled beams propagate through single-mode optical fibers (S630-HP Nufern) 21 m long each, which are extended from the source to two identically equipped stations. The fibers are inserted into flexible stainless steel tubes (12 mm inner diameter, 16 mm external), which traverse the lab's walls through drilled holes to the adjacent corridor and are placed over cable trays until reaching the stations. Each station's optics and electronics are mounted on a small wheeled optical table. Optics is placed inside a black box that protects it from spurious light and dust. The stainless steel tubes reach into these boxes. When measuring at short $L$, the stations are moved inside the lab and the tubes are bent to re-enter the lab through its door. Excepting for the distance in straight line between the stations, observations in the corridor or within the lab have identical values of all experimental parameters.

In each station, "bat-ears" are used to compensate birefringence distortion in the fibers. Polarization is measured with two exit fiber optic analyzers (Thorlabs PFC-780SM-FC). An auxiliary laser beam at 808 nm is inserted into the fibers at the source through removable mirrors EF (see Fig.SM1). The bat-ears are adjusted to maximize contrast between the two output gates of the fiber polarizers. For this action the SPCM are replaced by standard photodiodes. In order to facilitate measuring the contrast, a half waveplate (HWP3) in a fast (about 100 Hz) rotating mount is inserted between the auxiliary laser and the fiberports optics, see Section 1.3 below. Transmission from the input of the focusing fiberports optics (PAF, see Figure SM1) to the detectors is measured 83% for Alice and 82% for Bob.

The beams leaving the two outputs of the fiber polarizers are sent to single photon counting modules (SPCM, AQR-13 and AQRH-13, from Perkin-Elmer-Pacer-Excelitas). These modules emit one TTL signal for each detected photon. The TTL signals are sent to channels #1 (outcome "1") and #2 (outcome "0") of the TDCs in each station. Detections' time values are stored. The TDCs have 10 ps nominal time resolution, but accuracy is reduced to 2 ns because of detectors' jitter. One PC in each station controls the duration of the observation run, the opening of files and their naming and saving, following the instructions sent via wifi by a "Mother" PC placed near the source.

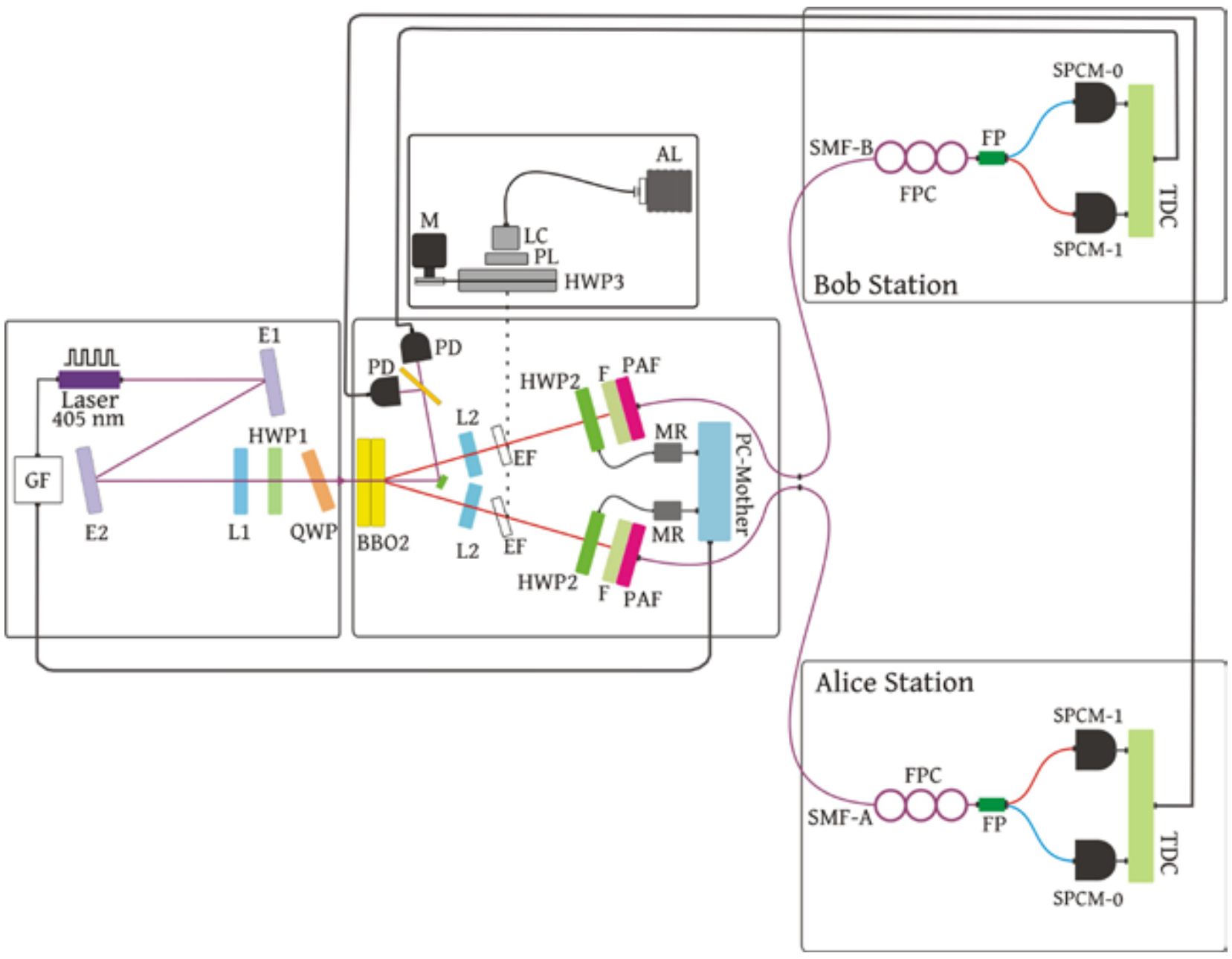

Figure SM1: Diagram of the setup. GF: function generator that modulates the laser repetition rate; L1,L2: f= 300 mm lenses; E1, E2: HR plane mirrors at 405 nm; HWP1 and QWP: half and quarter waveplates at 405 nm; BBO2: crossed BBO-I crystals (source of entangled states); PD: fast photodiodes, they send trigger signals of pulses' emissions to the TDCs via coaxial 50 Ω cables; HWP2, HWP3: half-waveplates at 810 nm; F: Interferential filters at 810 nm, $\Delta\lambda$=10 nm, 90% transmission; EF: auxiliary, removable HR plane mirrors in flip-flop mountings; AL: auxiliary CW laser diode at 810 nm coupled to a multi-mode fiber; LC: collimating optics; PL: linear polarizer; M: motor that rotates HWP3; MR: servo motor controllers of HWP2; PAF: fiberports f=7.5mm; SMF-A and B: single-mode fiber coils, 21 m total length each; FPC: birefringence compensator ("bat-ears"); FP: fiber polarization analyzer; SPCM: photon counting module; TDC: time-to-digital converters. The stations can be placed at different distances in straight line, what is a unique feature of this setup.

“Mother” directly controls the function generator that pulses the laser (including the switching or modulation of the repetition rate, following a previously specified program) [43] and the servo motors that adjust the settings angles. She also controls remotely wifi through a TCP/IP communication via a local network, the “sons” PCs in each station (Alice and Bob) to open, name and close the data files recorded in each TDC. Raw data are saved in .bin format. For a single 30 s run they require typically 300 kbit for each file of photons’ detections times, and 120 Mbit for the file of pulses’ detection times. After being processed and summed up, data of a whole session (typically ≈200 runs) require less than 20 Mbit in .dat format. We are eager to share our raw and/or processed data upon reasonable request.

The coaxial cables carrying the pulses’ start signal are 38 m long each, but the fibers are only 21 m long each. This means that “trigger” signals arrive to the TDCs later than “signal” photons. This is not a problem, for the TDCs record data continuously in all the input channels. The delay is observed to be constant (≈57 ns) and is taken into account for data processing. Photons’ detection times are positioned in reference to the trigger. Note that, in each station, time values in the three channels are measured by a single clock. Synchronization between the clocks in each station is achieved through the trigger pulses arriving to channels #3, and modulation of the pulses’ frequency, as explained.

*1.2. Procedure of data recording.*

We call one “run” an interrupted period of recording data. Data are saved in the PCs at the end of each run. Once the controlling program is started, the setup is able to perform an arbitrary number of successive runs with different settings (which are previously specified in a .txt file in Mother) without the operators’ assistance.

Typically, each run records data during 30s of real time. Recording runs are gathered in sets named “experiments”, which accumulate the results of 34 runs, scanning a complete set of angle values {α,β}. Because of idle periods of time inserted to give time the PCs to download from the TDCs and save the data files reliably, each “experiment” lasts about one hour. Typical curves of coincidences obtained in one “experiment” are shown in Figure SM2.

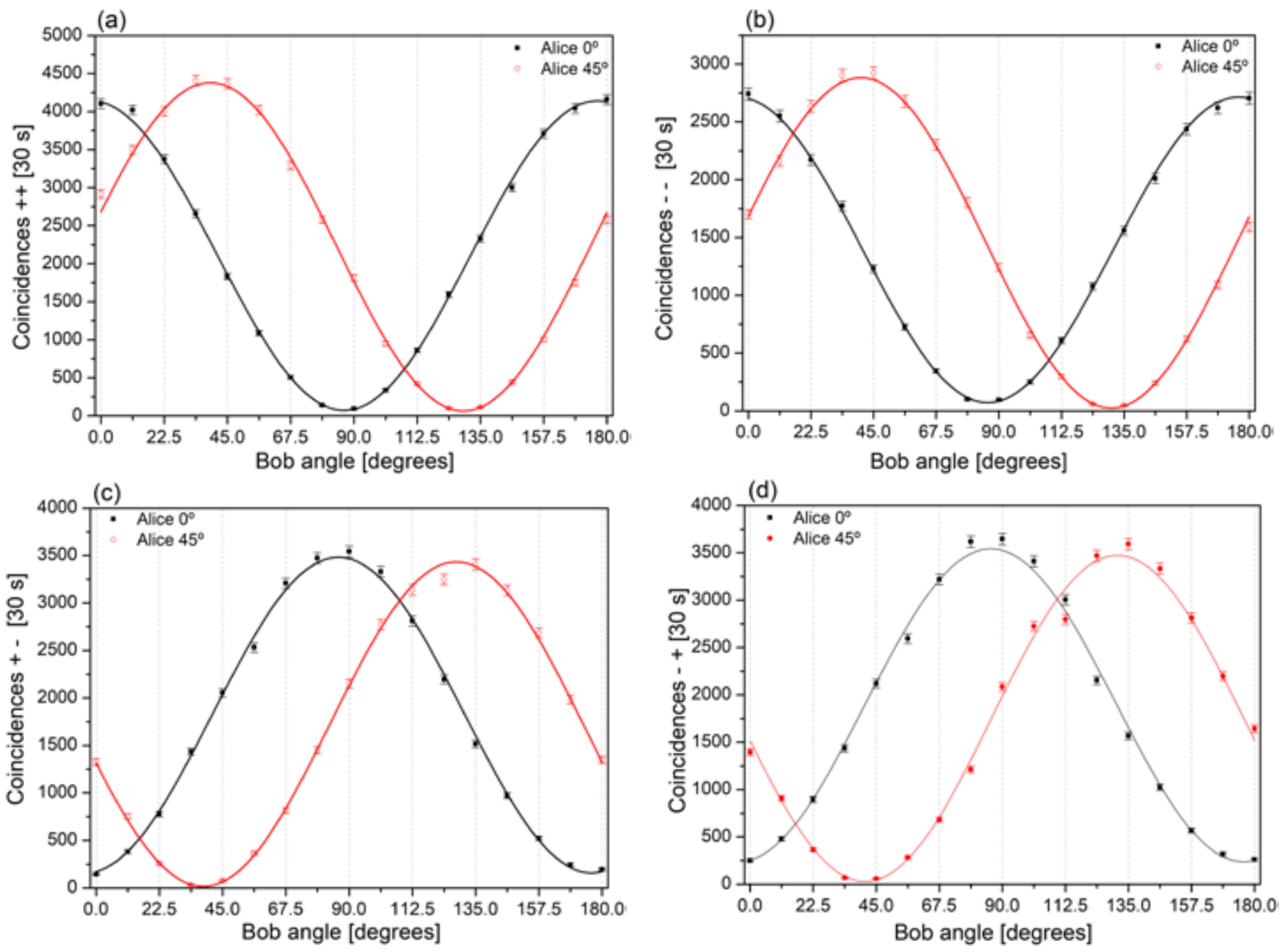

Figure SM2: Illustration of the curves recorded in one “experiment”; (a) Total number of coincidences as a function of the setting angles for “+,+” coincidences (i.e., coincidences between the detectors “+” in each station), (b) for “-,-”, (c) for “+,-”, (d) for “-,+”. In this case, measured $S_{CHSH}$ = 2.75, $L$=24 m.

A first checking experiment is performed, and sets of curves of coincidences as function of {α,β} are drawn. If the obtained curves are not satisfactory, realignment and/or improved birefringence compensation are performed. The checking experiment is then repeated. If everything is satisfactory instead, several further experiments are carried out until sufficient statistics is accumulated. Typically, 4 valid experiments are performed during each session.

Photons' detections times in channels #1 and #2 are positioned in reference to the trigger signals in channel #3. After summing up data produced by millions of pulses ($\approx 5\times10^8$ in a single experiment), plots of Singles and Coincidences as a function of time are obtained with satisfactory statistics. Frequency modulation or switch of the pulses' frequency determines the numbering of each pumping pulse to be the same in each station. Data processing shows that detections occurring during pulses with the same numbering are coincident within 2 ns. There are practically no coincident detections observed outside the pump pulses, which agrees with the following estimation: for a 2 ns coincidence time window and detectors' typical dark count rate of 200 $s^{-1}$, the number of accidental coincidences accumulated during a 30 s run is: $(200\ s^{-1})^2 \times 2.10^{-9}\ s \times 30\ s \approx 2.5\times10^{-3}$ in each time slot. Therefore, even if many runs are accumulated ($\approx$200 during one session of 6 experiments), only rarely one coincidence occurs in a slot outside the pulses.

As an illustration, typical time variations of singles and coincidences for one of the detectors (A+) are displayed in Figure SM3. These curves match the laser pump pulse shape as observed with a fast photodiode. The small fluctuations from the ideal pulse shape do not appear in $S_{CHSH}(t)$, which is remarkably square (see Fig.3 in the main text). We had observed this effect before [36,45].

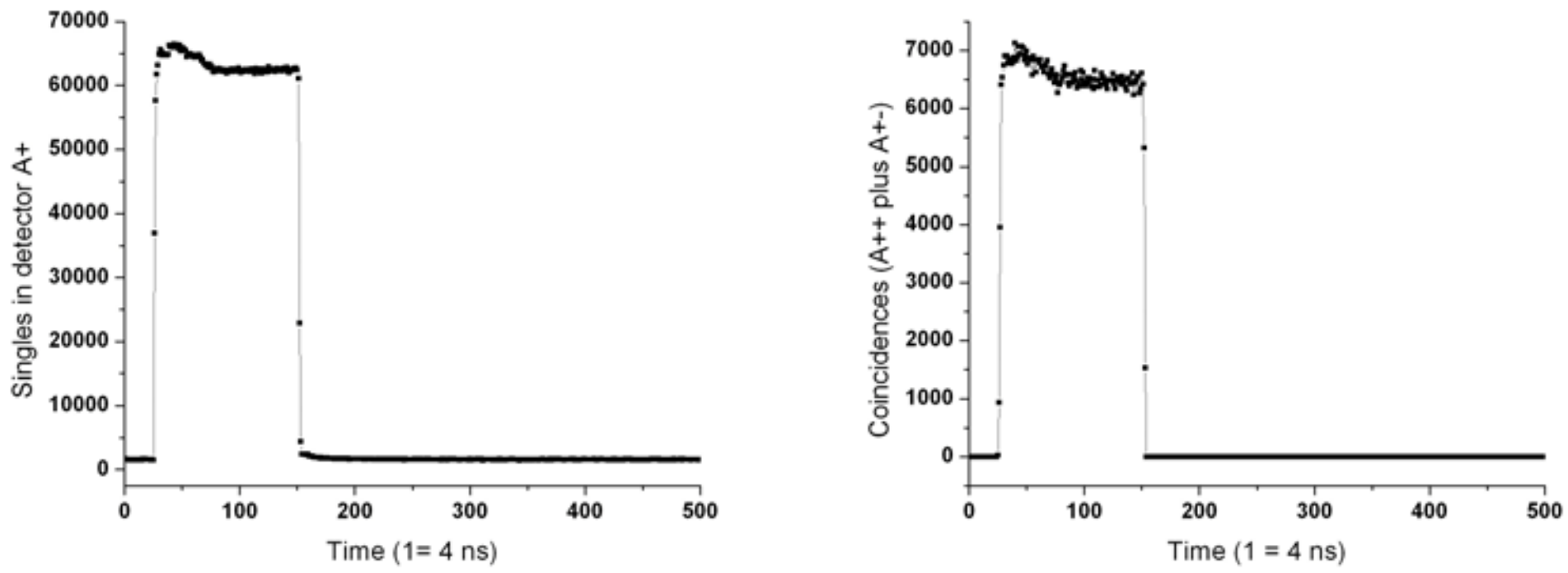

Figure SM3: Stroboscopic reconstruction of time evolution of singles and coincidences in detector "+" in station Alice (A+). Left: single detections, Right: coincidences (A++ plus A+-). Coincidences outside the pump pulses are zero, as expected. The number of singles outside the pump pulses ($\approx$1600 in each 4 ns slot) is consistent with the measured rate of dark counts of detector A+ (140 $s^{-1}$).

*1.3. Birefringence compensation.*

In order to make easier the (mostly try-and-error) method to compensate birefringence in the fibers with bat-ears, we use an auxiliary laser diode at 810 nm. This is practically the same wavelength of the entangled photons. The laser is fiber coupled (multi-mode). The beam is collimated, polarized and passed through a half wavelength waveplate. This waveplate is in a rotating mounting driven by a motor at ≈50 Hz. The result is a polarized beam at 810 nm which plane of polarization rotates at ≈100 Hz. This beam is inserted into the single mode fibers by using high reflection (at $45^o$) plane mirrors on flip-flop mountings (so that the mirrors can be easily removed from the optical axes, see Fig.SM1).

In each station, two photodiodes are placed at the exits of the fiber polarizer, and their outputs observed in an oscilloscope, see Figure SM4. The first coil in the bat ear (a quarter waveplate) is adjusted to maximize contrast. This means the point (representing the laser field) in the Poincaré sphere to be in the equator. Then the motor is stopped, and the position of the rotating waveplate is manually adjusted to the position that corresponds to one of the output fibers in the fiber polarizer (this position was previously marked on the mounting within which the waveplate rotates). Then the half waveplate in the bat ear is adjusted to maximize the signal (which is constant in time now) in one of the photodiodes, and to minimize the signal in the other. Then the waveplate is put into rotation again, and fine adjustments are introduced. The free coil in the bat-ear (which is theoretically unnecessary) is used, in practice, to adjust contrast in the diagonal basis. The procedure is repeated at the other station.

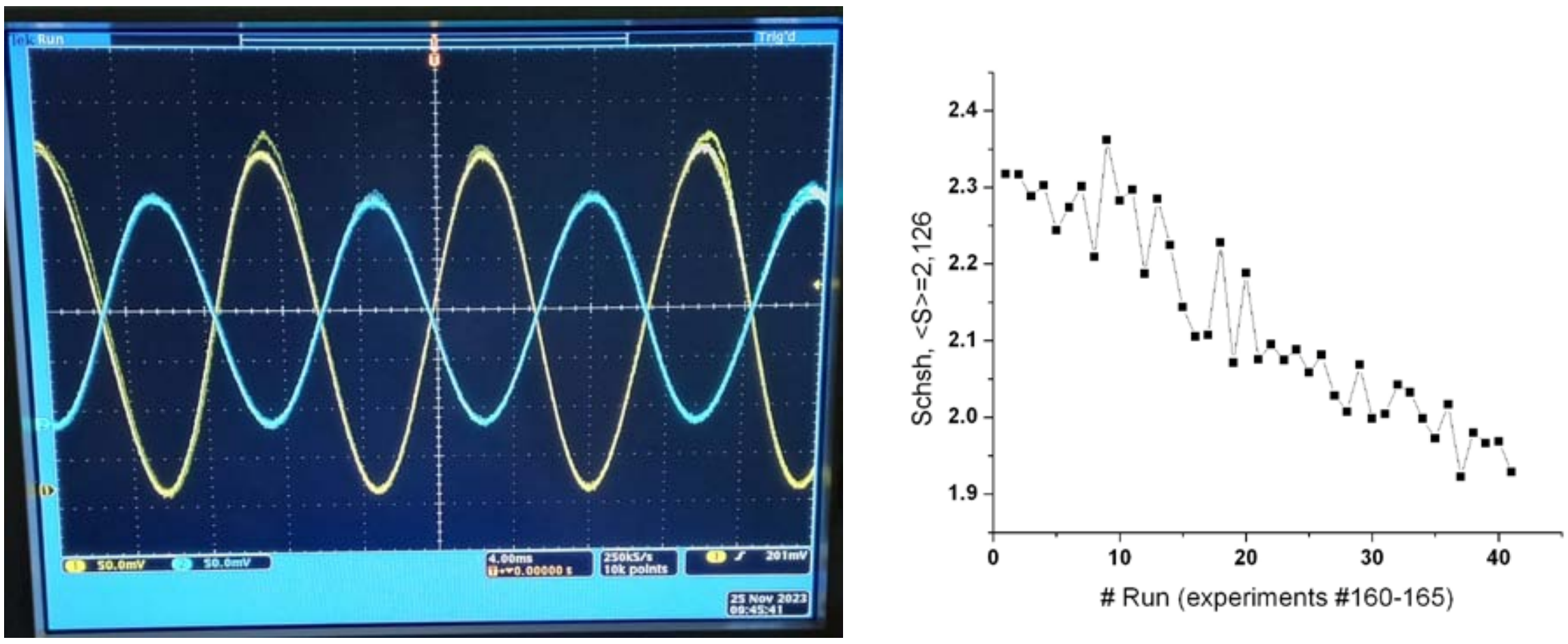

Figure SM4: Left: appearance of the photodiodes' signals when the auxiliary waveplate is rotating, the "bat-ears" are adjusted to maximize contrast of the (opposite phase) sinusoids. Right: example of the decay of $S_{CHSH}$ along the day (about 10 hs) in a condition of deficient birefringence compensation.

Thermal and mechanical perturbations affect birefringence during the day. In Fig.SM4, the decay of the value of $S_{CHSH}$ with real time is shown along the day (November 1$^{st}$, 2023), during a period of about 10 hs. This puts a practical limit to the total duration of a measuring session without new adjustments. Careful birefringence compensation using two photodiodes as described above is proven to be important to get starting values good enough to allow recording data during a whole day without needing further adjustments.

*1.4 Justification of assumptions.*

An ideal experiment to reveal the false hypothesis, as it is proposed in the main text, is unattainable nowadays. Due to detectors' low efficiency (even cryogenic nanowires), loophole-free violation of the BI can be reached with photons only by using Eberhardt's states, which unavoidably produce strongly non-uniform series [40]. Extractors of randomness are out of the question because they would burden the trend in randomness variation we intend to detect. Setups exploiting entanglement swapping between photons and matter do use Bell states (which, in controlled conditions, can produce uniform series), but they currently have a rate of detections too low to be suitable.

A simple solution at hand is to accept the *fair sampling* assumption valid [2]. It means that the set of recorded coincidences is an unbiased statistical sample of the whole set of detected and non-detected photons. In other words, that there are no conspiratorial mechanisms of photons' detection. The performed loophole-free experiments demonstrated these mechanisms to be inexistent or irrelevant when the experimenters' aim was to test the violation of BI. It seems unreasonable speculating these mechanisms exist and conspire to hide variations of randomness in our case. Otherwise, it would mean speculating that the setup knows the observers' *purposes* when performing an experiment! Nevertheless, reasonable or not, fair sampling does mean an additional assumption. Under this assumption, Bell states and existing single-photon detectors can be used.

Another problem is to achieve fast and unpredictable setting changes, necessary to close the predictability or measuring-independence loophole. In addition to the technical difficulty, there is the logical problem of performing unpredictable setting changes. An intrinsically untestable hypothesis is always involved. For, the changes must be *unpredictable by the source* of entangled states, but *not by the observer*. Both problems can be circumvented by assuming that any conspiratorial correlation between the stations vanishes when the source of entangled states is turned off, what occurs during the dead time between the pump pulses. The validity of this assumption is supported by the following observation: in a pulsed Bell's setup, the measured $S_{CHSH}$ parameter decays following a definite curve if the time coincidence window is increased beyond the pulse duration (see Figure SM5). This curve fits the one predicted assuming that detections outside the pulse are fully uncorrelated [45]. Assuming non-correlation implies the curve but, of course, observing the curve does not necessarily imply non-correlation. Some unknown conspiratorial effect might reproduce the curve, even if the physical systems at the stations remain correlated in some "hidden" way. Nevertheless, if one accepts that observing the curve implies non-correlation of the systems at the stations (which is the most reasonable choice), then unpredictable settings' changes are unnecessary.

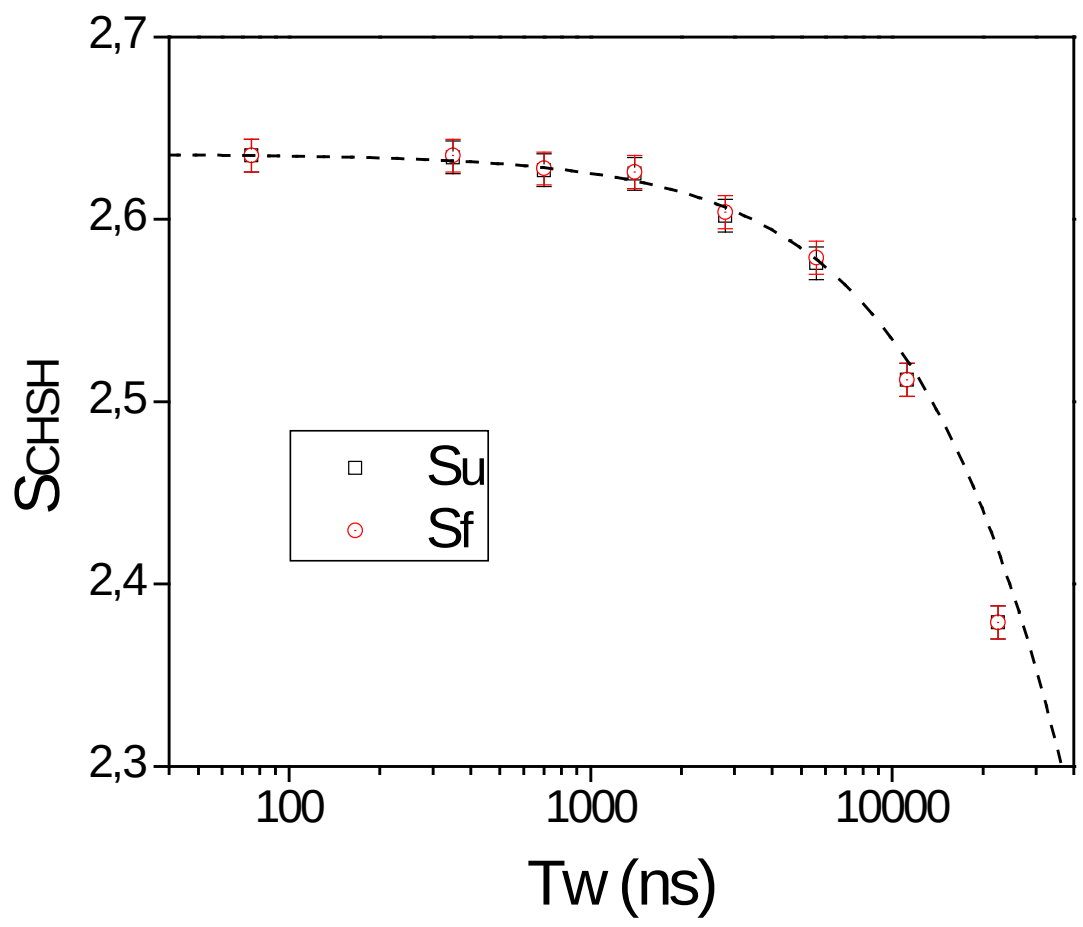

Figure SM5: Variation of $S_{CHSH}$ vs size of coincidence time window $T_w$ (from [45]). Measured values fit the curve obtained by assuming that all coincidences outside the pulse are fully uncorrelated. This result supports the second of the assumptions made in our experiment.

*Under these two assumptions* ("fair sampling" and "uncorrelated when source is turned off") the experiment is feasible, even with limited means.

## 2. Relationship between falsity of the main hypotheses and series' randomness (review).

### *2.1 Locality.*

According to the idea of Quantum Certified Randomness (QCR), the binary series of outcomes produced in the Fig.1 in the main text are *intrinsically* random. As "random" is a feature difficult to define (see Section 3 below), this idea is most appealing. The setup in Fig.1 would provide not only series to be used in practice, but also a definition: *random series is what is produced by this setup.*

QCR is supported by three main arguments:

*i)* Because of a numerical relationship, entanglement parameter $S_{CHSH}$ puts a lower bound to the series' minimum entropy *Hm* [49]:

$$Hm \equiv -\log_2 [max_r\, P(r)] \geq 1 - \log_2\{1+(2-S_{CHSH}^2/4)^{½}\} \qquad (2.1.1)$$

where P(*r*) is the probability of obtaining the outcome *r* in the series. If $S_{CHSH}$ reaches its maximum value $2\sqrt{2}$, then *Hm* = 1, which is the value of an ideally uniform series. An ideal random series has *Hm*=1 and also (because of its very definition) "one random bit per bit", but the inverse is *not* true.

F.ex., the series of digits of Champernowne's number or $\pi$ are demonstrated to have *Hm*=1 but are generated by algorithms. They are predictable and hence, not random (for all definitions of "random"). *Hm* is a good measure of the series' uniformity. Uniformity is a condition necessary, *but not sufficient*, for randomness.

By the way, in the experiment reported here, for $L$ = 24m, $S_{CHSH}$ = 2.73 and from eq.2.1.1 then $Hm \geq 0.55$; for $L$ = 1.5m, $S_{CHSH}$ = 2.62 and hence $Hm \geq 0.38$. The series in this report all have $Hm >$ 0.69, well above those bounds.

*ii)* An argument based on the Kochen-Specker theorem claims that the outcomes of some quantum experiments cannot be assigned by a program running on a classical Turing machine [50]. That is, they are *Turing non-computable*.

*iii)* If the existence of *non*-Local effects is taken as an axiom (i.e., if Locality is false) then the series of outcomes produced by space-like separated measurements (on a spatially spread entangled state) cannot be predicted by any method. Otherwise, faster than light signaling would be possible [33].

The argument *(i)* above guarantees a minimum level of uniformity of the series but, as said, a series can be ideally uniform and still be predictable (and hence, not random). Regarding *(ii)*, a series can be Turing non-computable and still not algorithmically random. Non-computability is a necessary, but not sufficient "symptom" of "true randomness" [51,52]. The argument (*iii)* is the strongest: it ensures that no algorithm can predict the series. We consider (*iii*) the only condition that certifies the series' randomness.

In short: if BI are violated because Locality is false (i.e., non-Local effects do exist), then the series of outcomes produced in space-like separated conditions in Fig.1 *must* be random. On the other hand, series of outcomes made of not-space-like separated detections are *not necessarily* random, even if they violate the BI [33].

*2.2 Ergodicity.*

Let first review why the hypothesis of Ergodicity is necessary in the *observation* (not in the derivation) of BI. F.ex. in the derivation of the CHSH bound, an intermediate step requires defining the following quantity [2]:

$$AB(\alpha,\beta,\lambda)= \frac{\{(+1).[C^{++} + C^{--}] + (-1)[C^{-+} + C^{+-}]\}}{\{C^{++} + C^{--} + C^{-+} + C^{+-}\}} \mid \alpha,\beta,\lambda \qquad (2.2.1)$$

where the $C^{ij}|_{\alpha,\beta}$ are the number of coincidences observed at detectors $\{i, j\}$ when the analyzers are set to values $\alpha,\beta$; $\lambda$ is the (unobservable) hidden variable. One does not know the values taken by $\lambda$; one can only observe the average over the $\lambda$-space; this is the crucial correlation parameter $E(\alpha,\beta)$:

$$E(\alpha,\beta) = \int d\lambda.\rho(\lambda).AB(\alpha,\beta,\lambda) \qquad (2.2.2)$$

The next expression in the derivation of the CHSH bound involves correlation parameters for different settings:

$$E(\alpha,\beta) - E(\alpha,\beta') = \int d\lambda.\rho(\lambda).AB(\alpha,\beta,\lambda) - \int d\lambda.\rho(\lambda).AB(\alpha,\beta',\lambda) = \int d\lambda.\rho(\lambda).[AB(\alpha,\beta,\lambda) - AB(\alpha,\beta',\lambda)] \qquad (2.2.3)$$

But there is a problem in the apparently innocent second equality. Actual measurements occur during time, and it is impossible measuring with two different settings ($\beta$ and $\beta$') at the same time. Let suppose the analyzer setting in Bob's station is $\beta$ from time $t=T/4$ to $t=T/2$, and $\beta$' from $t$=0 to $t=T/4$, then:

$$E(\alpha,\beta)-E(\alpha,\beta')=(4/T)\int_{T/4}^{T/2}dt.\rho(t).AB(\alpha,\beta,t)-(4/T)\int_{0}^{T/4}dt.\rho(t).AB(\alpha,\beta',t)$$
$$\neq(1/T)\int_{0}^{T}dt.\rho(t).[AB(\alpha,\beta,t)-AB(\alpha,\beta',t)] \qquad (2.2.4)$$

The rhs in the first line is what is actually measured, while the integral in the second line is the expression usually considered equivalent to the last term in eq.2.2.3, which leads to the CHSH bound. The reason of the difference in eq.2.2.4 is that the integration intervals are different. This cannot be solved, because it is impossible measuring with settings $\beta$ and $\beta$' at the *same* time. In order to retrieve the usual CHSH bound, one must assume, in addition to Locality and Realism, that:

$$\int d\lambda.\rho(\lambda).AB(\alpha,\beta,\lambda)=(1/\Delta T)\int_{\theta}^{\theta+\Delta T}dt.\rho(t).AB(\alpha,\beta,t) \qquad (2.2.5)$$

for all settings and time values $\theta$, where $\Delta T$ is a "sufficiently long" time. As the integral in the lhs is unique, all integrals in the rhs are equal regardless the value of $\theta$. This expression means the equality between an ensemble average (the integration over the hidden variables' space, lhs) and a time average (rhs); that's why V.Buonomano, who was the first to realize this limitation, appropriately named it "Ergodicity". Without this hypothesis, the observation of the violation of BI cannot be justified [22-27]. Strictly speaking, Ergodicity is not the most general possible assumption that allows retrieving the usual expression of the BI, but it is the most physically sounding.

Let us see now the relationship between Ergodicity and randomness.

It can be visualized as follows: a classical system that evolves ergodically explores its (bounded) phase space evenly, that's why time and ensemble averages are equal. If the evolution is non-ergodic instead, then the system spends more time in some regions of its phase space than in others of the same measure (for, time and ensemble averages are assumed not-equal). Hence, at a given time, the system is more probably found in some regions than in others. Its future state can be partially predicted. Therefore, an evolution that is non-ergodic is (at least partially) predictable and hence not-random, for all definitions of "random".

In more formal terms: let suppose that the outcomes in Fig.1 are caused by the evolution of an underlying dynamical system. Let partition the phase space of this system as follows: label "1"("0") the regions where the dynamics causes a "1"("0") in the series. Inside these regions, there are sub-regions where the dynamics causes the strings 11, 10 (01, 00). Following this partition up to strings of length $n$ ($n$ is large, but much smaller than the total length of the series), the phase space is divided into $2^n$ sub-regions. When the system evolves into one of these sub-regions, the corresponding string appears in the observed series.

Birkhoff's theorem ensures that Ergodicity is valid *if and only if* the phase space is metrically un-decomposable. Therefore, if Ergodicity is not valid, then the phase space is metrically decomposable. This means that it can be divided into two regions of measure different from 0 or 1 that are invariant during the system's evolution [34]. The system is then trapped into one of these regions, and it never enters into the other one. The invariant regions have measure different from 0 or 1, hence the non-visited region includes a finite number of the $2^n$ labeled sub-regions, which are never visited by the system. In consequence, there is a finite number of strings of length $n$ that do not appear in the complete series. The complete series is then, by definition, not uniform. As uniformity is a necessary condition for randomness, a non-Ergodic evolution (of the assumed underlying classical dynamical system) necessarily generates non-random series.

In summary: Random $\Rightarrow$ Ergodic, but we do not claim this relationship to be general (see below).

There are two examples in the web claiming to be systems that are random and not-ergodic, in contradiction with the implication just stated. They are both erroneous, as shown next:

*First example:* consider a random process where a particle moves in the unit square in two dimensions. Divide the unit square into four smaller squares, and label them 0 or 1 in the following way:

00
11

Now suppose that the particle spends 90% of its time in the left half of the square and only 10% in the right half, but that it randomly jumps between top and bottom. Then, it is claimed, the outcome series of zeros and ones is random, even though the particle motion is not ergodic.

The error in this example is that two different partitions of phase space are being mixed. If "left" and "right" are indistinguishable, then the series is random and the evolution is ergodic. If the columns are distinguishable instead, then the elements of the series are: $\{0_{left}, 0_{right}, 1_{left}, 1_{right}\}$. In order to retrieve a binary series, let code these elements as: {00, 01, 10, 11}. The resulting series is not uniform, for there are more strings 00 (or $0_{left}$) and 10 (or $1_{left}$) than 01 and 11. Hence, the series is *not* random (and the evolution is not ergodic). Both alternatives are compatible with the implication: Random ⇒ Ergodic.

*Second example:* Suppose two coins, one is fair and the other has two heads. One of the coins is randomly chosen first, and then a sequence of tosses is performed on the chosen coin. Then the ensemble average is ½.(½+1) = ¾; yet the long-term average is ½ for the fair coin and 1 for the two-headed coin. So the long term time-average is *either* ½ *or* 1. Hence, it is claimed, this random process is not ergodic in mean.

The error in this example is easily visible if the number of tosses of the chosen coin is made explicit. Let suppose that that number is 3, and that "1" ("0") corresponds to "head" ("tail"). Then a section of the *complete* series has the form:

111010100111111001000…

(the two-heads coin was chosen in the first, fourth and fifth occasion). Such a series is not uniform for there is an excess of the 111 string, hence, it is not random. Its *complete* time average is ¾ and coincides with the ensemble average. Therefore the process is ergodic, but the series is not random. Also this case is compatible with the implication Random ⇒ Ergodic.

Anyway, we claim the reasoning leading to Random ⇒ Ergodic to apply only to the case of interest here, that is, a classical dynamical system which produces binary series in a Bell's setup as it enters different regions of its *bounded* phase space. In this case, and in this case only, if the evolution of the system is not Ergodic, then the produced series is not uniform (and hence, not random). The relationship Random ⇒ Ergodic is not claimed to be general. F.ex., the unbounded random walk is random by definition, and evidently non-Ergodic.

### *2.3 Falsity of Realism.*

The Copenhagen interpretation of QM is the most important of the descriptions of BI's violation that hypothesizes Realism to be false. Strictly speaking, this interpretation says *nothing* about series' randomness. Born's rule allows calculating the probability of an outcome, but is silent about the features of the time series underlying the measurement of that probability. The only explicit opinion on this subject is von Neumann's axiom. It states that quantum measurements violate Leibniz's principle of sufficient reason. Therefore, the outcomes "1" or "0" in Fig.1 have no previous cause. A series of such outcomes is intuitively random, but this intuition is difficult to formalize [35]. Besides, von Neumann's axiom can be understood in two ways, or strengths. Its "strong" form means that Leibniz's principle is violated in quantum experiments. The "weak" form means that the axiom is part of a user's guide or warning about what QM formalism can or cannot predict, but not necessarily a feature to be experimentally observed.

In short: if BI are violated because Realism is false, then there is no reason to say the series produced in a Bell's experiment must be random, or not random. Several studies that apply standard tests of randomness to quantum generated series support this result [25,39,55-56].

## 3. Discussion about the meaning of "Randomness".

There is no universally accepted definition of randomness. We only know for sure that a predictable series (even partially predictable) is not-random. Nevertheless, there are two definitions that are relevant from a practical point of view:

*i)* A binary series is "statistically random", uniform or Borel normal if the number of strings of "1" and "0" of different length $n$ (say, 110101 for $n$=6), matches (statistically) the number in a series produced by tossing an ideal coin. Other tests of statistical randomness measure the decay of the self-correlation or the mutual information, or calculate entropies. They all involve probabilities and require, in principle, the series to be stationary (what requires further tests). The battery of tests provided by the National Institute of Standards and Technology (NIST) mostly checks this definition.

*ii)* A binary series is "algorithmically random" if there is no classical program code able to generate the series using a number of bits shorter than the series itself. Note that this definition does not involve probabilities, and applies even to non-stationary and short series. It is directly related with the idea of *complexity* developed by Chaitin and Kolmogorov [53]. In few words, the complexity of a series of length $N$ is the length $K$ of the shortest classical program able to generate the series. If $K \approx N$ the series is said to be algorithmically random or incompressible, which is often considered the strongest form of randomness. The problem is that $K$ cannot be computed, for one can never be sure there is no shorter program able to generate the series. It can be only estimated from the asymptotic compressibility of the series using, f.ex., the algorithm devised by Lempel and Ziv [38]. We call $Kc$ the estimated and normalized value of $K$. An algorithmically complex series is non-computable and Borel normal, but the inverses are not true [54]. In order to calculate $Kc$, we use the approach developed by Kaspar and Schuster [46] and implemented by Mihailovic [47].

The goal is to express the series $\{x(1),x(2),\ldots,x(N)\}$ using the smallest possible number of bits. This number defines the series' complexity. The algorithm by Lempel and Ziv adds a new "word" to its memory every time it finds a substring of consecutive elements not previously registered. The size of the compiled vocabulary, and the rate at which new words are found, are then the basic ingredients to estimate complexity. Then the complexity counter $c(N)$, which is defined as the minimum number of distinct words in a given sequence of length $N$, is calculated. It is demonstrated that, as $N \to \infty$, $c(N) \to N/log_2(N)$ in an ideally incompressible series. The *normalized complexity measure Kc* is then defined (for the complete series of length $N$) as:

$$Kc \equiv c(N) \times log_2(N)/N \tag{3.1}$$

In this way, the value of $Kc$ is near to 0 for a periodic or regular sequence, and near to 1 for an incompressible one.

Be aware that *Kc estimates* complexity, it does not *measure* it. The famous example is the series of digits of $\pi$, which is computed to have $Kc \to 0.95$ ($N \to \infty$). Yet, the series is generated by an algorithm that requires only a few bits. Its true normalized complexity converges to zero. On the other hand, if $Kc$ is low, then the series is certainly compressible. Once again, randomness (or complexity) cannot be demonstrated; only not-randomness can be demonstrated. Note that $Kc \geq K$.

## 4. Reconstructing an attractor from time series.

Takens' embedding theorem states that the structure of an underlying object in phase space can be revealed, *no matter what is the original phase space*, in a vector space of dimension dE reconstructed from a time series of scalar measurements $\{x(1),x(2),\ldots,x(N)\}$. This is done by building sets of *vectors* of dimension dE with the form [29]:

$$\{x(n), x(n+m), x(n+2m), \ldots, x(n+(\mathrm{dE}-1)m)\} \qquad (4.1)$$

The value of the delay *m* is usually chosen equal to the first minimum in the average mutual information calculated along the whole series. This choosing is not mandatory but is, in general, the most efficient one. For, the amount of information added by each new element in the vector is maximal (in the average). Results are checked against variations in the value of *m*.

The number of dimensions necessary to unfold the object in phase space defines the value of dE. That number is found by eliminating the apparent overlaps existing in a reconstruction like eq.4.1 of insufficient dimension. F.ex.: let suppose that the object in phase space is a cube. In a one-dimension reconstruction, a segment of the line is obtained. Many points of the cube, that are actually distant, appear close to each other within that segment of the line. In a two-dimension reconstruction instead, many of the former near neighboring points are revealed to be, in fact, distant. They are revealed to have been *false nearest neighbors* (fnn) in the one-dimension reconstruction. When going to a three-dimension reconstruction, more points are revealed to have been fnn, for the same reason. Yet, going beyond three dimensions does not reduce the number of fnn any longer, for the cube *is* an object in three dimensions. The unfolding is complete. That's how it is possible to know, from the time series, the value of dE.

Therefore, if a compact object in phase space does exist, the number of fnn decays to zero (or near to zero, for noisy series of experimental origin) *and remains there* as the number of dimensions of the reconstruction increases. The value of dE is then given by the first value of dimension for which fnn reaches zero. If no attractor exists instead, the number of fnn never reaches zero. Typically, it decreases to a minimum far from zero, and then rises again with increasing dE.

In practice, in order to rule out "false positives" in the search of attractors, the series is scrambled. This destroys the time order, that is, the dynamics producing the series (if it exists). If the fnn of the scrambled (or "surrogate") series still decays to near zero, it means that the decay is the consequence of a particular distribution of the series' values, but not a consequence of the dynamics. No attractor, in consequence, has been found. On the other hand, if the fnn of the surrogate series does not decay, then it is an indication that the original series is, in fact, produced by an underlying dynamics. In practice, it is advisable to check this feature to be robust against a different choosing of the value of the delay *m*.

There is an issue related with the finite length of the series. As dE increases, the volume available for the data concentrates in the periphery of the space, and it may happen that no points of the reconstructed object are near neighbors. To account for this, it is necessary to require that the distance added in going up one dimension to not to be larger than the size of the object (which is obtained as a statistical estimation of the dispersion of the set of points of the reconstruction). When the distance grows too much, the number of neighbors becomes scarce, and it is not possible to go further (in the value of dE). This situation is named "condition of insufficient data". This condition is fatally found if dE is increased too much. If this occurs before the number of fnn reaches zero (if the series has an underlying low dimensional cause), or a minimum and starts increasing again (if the series has not such cause) the nature of the series (that is: whether there is an underlying object in phase space, or not) cannot be determined.

In the cases a value of dE is reliably established, it is possible to calculate the Lyapunov exponents of the dynamics. A positive Lyapunov indicates the dynamics is chaotic. It means that the series becomes unpredictable (in the average) after a time given by the inverse of the largest Lyapunov exponent.

If the density of available points in phase space is large enough, it is possible to attempt measuring the dimension of the attractor dA ($\mathrm{dE} \leq 1 + 2\times\mathrm{dA}$). The value of dA may be fractal. However, density of points of experimental series is rarely sufficient to make this measurement reliably.

Finding an attractor in experimental time series is therefore like applying a series of pass / reject filters. The first filter is to observe the decay of fnn to zero. If this does not occur, then the series can be discarded. The second filter, in the cases that that decay does occur, is to eliminate false positives by observing the behavior of fnn in the surrogate series. The third filter is to check that the same results are obtained if the delay *m* is changed. A fourth filter is to check whether the attractor is caused by some dynamics of the pump laser. This is done by applying the three filters mentioned before to the corresponding files of *single* detections (there are two of them for each series of coincidences, one for each station). The original series passes the fourth filter if both series of singles do *not* pass the three filters.

Definitive proof of the presence of an attractor is to reconstruct it and to predict future elements in the series successfully (in the average), at least up to the horizon of predictability estimated from the inverse of the largest Lyapunov's exponent.

The TISEAN (free available) package is designed to calculate all these parameters, and to automatically take into account the limitations arising from scarce data and to warn the user about the results' reliability [55].

**5. Student's t-test and other tests.**

Student's t-test is performed to determine whether the means (of the 136 binary series in each slot) of *Kc* and *Hm* differ significantly along the pulses. In the case *L*=24m, the comparison between the condition of space-like separated generation of series (first slot) and the subsequent ones (not space-like separated) is of main interest. The case *L*=1.5m (not space-like separated in all slots) works as a reference. So we compare the results of the t-value of the mean in the first slot with the means in the subsequent slots. The critical value for 4 degrees of freedom and significance value $\alpha$=0.95 is $t_{critical}$ = 2.776, note that in all cases (see the tables below) t-value < $t_{critical}$. Besides, there is no definite trend in their variation, neither difference between the *L*=24 m case, where space-like separated detections occur, and the reference case *L*=1.5m.

We present the results for *Kc* and 5 slots in Table 1 first, the other cases follow.

| | *Kc* t-value<br>Alice station (*L* = 1.5 m) | *Kc* t-value<br>Bob station (*L* =1.5m) | *Kc* t-value<br>Alice station (*L* =24m) | *Kc* t-value<br>Bob station (*L* =24m) |
|---|---|---|---|---|
| Slot 2 | 0.28 | 0.28 | 0.20 | 0.03 |
| Slot 3 | 0.33 | 0.31 | 0.24 | 0.09 |
| Slot 4 | 0.15 | 0.07 | 0.15 | 0.21 |
| Slot 5 | 0.45 | 0.44 | 0.34 | 0.15 |

Table 1: t-values comparing the means of estimated complexity of the first slot with the subsequent ones, 5 slots (4 degrees of freedom), $t_{critical}$ = 2.776 for significance value $\alpha$=0.95.

| | *Hm* t-value<br>Alice station (*L* =1.5m) | *Hm* t-value<br>Bob station (*L* =1.5m) | *Hm* t-value<br>Alice station (*L* =24m) | *Hm* t-value<br>Bob station (*L* =24m) |
|---|---|---|---|---|
| Slot 2 | 0.14 | 0.07 | 0.04 | 0.07 |
| Slot 3 | 0.31 | 0.12 | 0.06 | 0.03 |
| Slot 4 | 0.33 | 0.19 | 0.09 | 0.04 |
| Slot 5 | 0.33 | 0.15 | 0.11 | 0.05 |

Table 2 (above): the same as Table 1, but for Minimum Entropy, $t_{critical}$ = 2.776 for significance value $\alpha$=0.95.

| | *Kc* t-value<br>Alice station (*L* =1.5 m) | *Kc* t-value<br>Bob station (*L* =1.5 m) | *Kc* t-value<br>Alice station (*L* =24m) | *Kc* t-value<br>Bob station (*L* =24m) |
|---|---|---|---|---|
| Slot 2 | 0.87 | 0.81 | 0.20 | 0.27 |
| Slot 3 | 0.53 | 0.45 | 0.23 | 0.34 |
| Slot 4 | 0.56 | 0.53 | 0.22 | 0.29 |
| Slot 5 | 0.69 | 0.68 | 0.30 | 0.37 |
| Slot 6 | 0.46 | 0.47 | 0.30 | 0.35 |
| Slot 7 | 0.27 | 0.23 | 0.01 | 0.02 |
| Slot 8 | 0.20 | 0.36 | 0.04 | 0.09 |
| Slot 9 | 0.03 | 0.06 | 0.16 | 0.16 |
| Slot 10 | 0.01 | 0.05 | 0.20 | 0.09 |

Table 3 (above): t-values comparing the means of estimated complexity of the first slot with the subsequent ones, 10 slots (9 degrees of freedom), $t_{critical}$ = 2.262 for significance value $\alpha$=0.95.

| | *Hm* t-value Alice station (*L* =1.5 m) | *Hm* t-value Bob station (*L* =1.5 m) | *Hm* t-value Alice station (*L* =24m) | *Hm* t-value Bob station (*L* =24m) |
|---|---|---|---|---|
| Slot 2 | 0.14 | 0.10 | 0.09 | 0.11 |
| Slot 3 | 0.11 | 0.05 | 0.10 | 0.15 |
| Slot 4 | 0.19 | 0.07 | 0.09 | 0.09 |
| Slot 5 | 0.19 | 0.06 | 0.03 | 0.06 |
| Slot 6 | 0.22 | 0.06 | 0.02 | 0.03 |
| Slot 7 | 0.14 | 0.08 | 0.02 | 0.02 |
| Slot 8 | 0.32 | 0.18 | 0.04 | 0.02 |
| Slot 9 | 0.07 | 0.02 | 0.07 | 0.04 |
| Slot 10 | 0.36 | 0.16 | 0.03 | 0.06 |

Table 4 (above): the same as Table 3, but for Minimum Entropy, $t_{critical}$ = 2.262 for significance value $\alpha$=0.95.

All t-values are much smaller than the critical value for significance value of 0.95. This means there is no significant statistical difference between the averages in each of the slots and the reference averages. In other words, (up to 95% statistical reliability) there is no trend during the pulse, or relationship between the slot number and the observed value. The values in the first slots (*L*=24 m) are not only well below $t_{critical}$, but also numerically close to the other values in the table.

We are concerned on the statistical significance of the apparent difference between the *Hm* curves when *L*=24m and *L*=1.5m. So we compare the t-values for the two cases (10 slots only). Now we have 19 degrees of freedom and for a significance value of 95%, $t_{critical}$ = 2.093.

| | *Hm*, Alice | *Hm*, Bob |
|---|---|---|
| Slot 1 *L*=1.5m, t-values ≤ | 0.83 | 0.84 |
| Slot 2 *L*=1.5m, t-values ≤ | 0.90 | 0.90 |
| Slot 3 *L*=1.5m, t-values ≤ | 0.88 | 0.87 |
| Slot 4 *L*=1.5m, t-values ≤ | 0.70 | 0.75 |
| Slot 5 *L*=1.5m, t-values ≤ | 0.71 | 0.76 |
| Slot 6 *L*=1.5m, t-values ≤ | 0.68 | 0.78 |
| Slot 7 *L*=1.5m, t-values ≤ | 0.73 | 0.75 |
| Slot 8 *L*=1.5m, t-values ≤ | 0.63 | 0.65 |
| Slot 9 *L*=1.5m, t-values ≤ | 0.79 | 0.81 |
| Slot 10 *L*=1.5m, t-values ≤ | 0.62 | 0.68 |
| Slot 1 *L*=24m, t-values ≤ | 0.85 | 0.71 |
| Slot 2 *L*=24m, t-values ≤ | 0.89 | 0.85 |
| Slot 3 *L*=24m, t-values ≤ | 0.90 | 0.90 |
| Slot 4 *L*=24m, t-values ≤ | 0.89 | 0.80 |
| Slot 5 *L*=24m, t-values ≤ | 0.82 | 0.78 |
| Slot 6 *L*=24m, t-values ≤ | 0.87 | 0.72 |
| Slot 7 *L*=24m, t-values ≤ | 0.84 | 0.73 |
| Slot 8 *L*=24m, t-values ≤ | 0.85 | 0.74 |
| Slot 9 *L*=24m, t-values ≤ | 0.80 | 0.71 |
| Slot 10 *L*=24m, t-values ≤ | 0.82 | 0.65 |

Table 5 (above): t-values for Minimum Entropy, 10 slots, comparison of the total set for cases *L*=1.5 m and *L*=24 m against each slot, $t_{critical}$ = 2.093 for 19 degrees of freedom and $\alpha$=0.95.

Once again, all t-values are well below $t_{critical}$ = 2.093, and besides, there is no perceivable difference if the slot being compared belongs to the *L*=1.5m or the *L*=24m case.

Finally, we consider the following questions:

1) Is the level of randomness when *L*=24m significantly different (i.e., beyond statistical fluctuations) from the level when *L*=1.5 m? Are the slopes different?

2) Are the values of *Hm* and *Kc* in the firsts slots when *L*=24 m significantly different from the values in the remaining slots? How this difference (if it exists) compares with the case when *L*=1.5 m?

These questions are answered, for each station and estimator of randomness, and for 10 slots, in Table 6 below. These answers are obtained from processing the data in the tables in the next section, which display the numerical values corresponding to Figs.5 and 6 in the main text.

| | **Alice**, *Hm* | **Alice**, *Kc* | **Bob**, *Hm* | **Bob**, *Kc* |
|---|---|---|---|---|
| Slots 1 and 3 overlap? (*L*=24 m) | Yes | Yes | Yes | Yes |
| Slot 1 and average over slots 3-10 overlap? (*L*=24 m) | Yes | Yes | Yes | Yes |
| Slope (linear fit) at slots 1-3 and slope at slots 3-10 overlap?, *L*= 24 m | Yes | Yes | Yes | Yes |
| Slope (linear fit) at slots 1-3 and slope at slots 3-10 overlap?, *L*= 1.5 m | Yes | Yes | Yes | Yes |
| Slope all slots *L*=24 m and *L*=1.5 m overlap? | Yes | Yes | Yes | Yes |

Table 6 (above): Answers to main questions, 10 slots.

All answers are "Yes", so that we find no perceivable difference between the *L*=24 m case and the reference case *L*=1.5 m.

We conclude that our data do not show variation of *Hm* or *Kc* (with statistical significance >95%) along the pulse, neither between the *L*=24m and the *L*=1.5m cases. We then infer there is no difference of "actual" randomness between the series made of space-like separated detections, and the series made of detections occurring inside the same light cone. This supports the idea that Realism is the false hypothesis.

## 6. Tables of data.

The value in each box in the Tables 7-14 below is obtained from averaging the results of 136 series. These series have a length ≈6 kbit each for the case of 5 slots, and ≈3 kbit for the case of 10 slots. Be aware that, because of the way they are composed, these series are different: the 3kbit ones are *not* merely the first or second half of the 6 kbit ones. Statistical dispersions are indicated. The last column shows the relative variation between the *L*=24m and *L*=1.5m cases. The variation is always small, and there is no recognizable trend.

We are eager to share our raw time stamped data under reasonable request.

| **ALICE station** | ***Kc, L= 24m*** | ***Kc, L= 1.5m*** | $Kc_{(L=24m)}/Kc_{(L=1.5m)}$ |
|---|---|---|---|
| Slot 1 | 0.881 ± 0.011 | 0.885 ± 0.010 | 0.996 |
| Slot 2 | 0.884 ± 0.012 | 0.880 ± 0.012 | 1.005 |
| Slot 3 | 0.885 ± 0.012 | 0.879 ± 0.014 | 1.007 |
| Slot 4 | 0.879 ± 0.012 | 0.887 ± 0.013 | 0.990 |
| Slot 5 | 0.876 ± 0.011 | 0.892 ± 0.014 | 0.981 |

Table 7 (above): Normalized estimated complexity, Alice station, 5 slots.

| **BOB station** | ***Kc, L= 24m*** | ***Kc, L= 1.5m*** | $Kc_{(L=24m)}/Kc_{(L=1.5m)}$ |
|---|---|---|---|
| Slot 1 | 0.885 ± 0.011 | 0.887 ± 0.010 | 0.998 |
| Slot 2 | 0.886 ± 0.012 | 0.883 ± 0.013 | 1.003 |
| Slot 3 | 0.887 ± 0.011 | 0.882 ± 0.014 | 1.005 |
| Slot 4 | 0.882 ± 0.011 | 0.889 ± 0.014 | 0.993 |
| Slot 5 | 0.883 ± 0.010 | 0.894 ± 0.013 | 0.987 |

Table 8 (above): Normalized estimated complexity, Bob station, 5 slots.

| ALICE station | *Hm, L= 24m* | *Hm, L= 1.5m* | $Hm_{(L=24m)}/Hm_{(L=1.5m)}$ |
|---|---|---|---|
| Slot 1 | 0.779 ± 0.062 | 0.841 ± 0.030 | 0.927 |
| Slot 2 | 0.775 ± 0.066 | 0.835 ± 0.028 | 0.929 |
| Slot 3 | 0.784 ± 0.060 | 0.828 ± 0.030 | 0.948 |
| Slot 4 | 0.786 ± 0.056 | 0.827 ± 0.029 | 0.951 |
| Slot 5 | 0.788 ± 0.057 | 0.827 ± 0.027 | 0.952 |

Table 9 (above): Minimum Entropy, Alice station, 5 slots.

| BOB station | *Hm, L= 24m* | *Hm, L= 1.5m* | $Hm_{(L=24m)}/Hm_{(L=1.5m)}$ |
|---|---|---|---|
| Slot 1 | 0.825 ± 0.064 | 0.875 ± 0.044 | 0.942 |
| Slot 2 | 0.819 ± 0.065 | 0.871 ± 0.044 | 0.941 |
| Slot 3 | 0.827 ± 0.061 | 0.868 ± 0.045 | 0.953 |
| Slot 4 | 0.829 ± 0.060 | 0.863 ± 0.048 | 0.960 |
| Slot 5 | 0.830 ± 0.061 | 0.866 ± 0.045 | 0.958 |

Table 10 (above): Minimum Entropy, Bob station, 5 slots.

| ALICE station | *Kc, L= 24m* | *Kc, L= 1.5m* | $Kc_{(L=24m)}/Kc_{(L=1.5m)}$ |
|---|---|---|---|
| Slot 1 | 0,887 ± 0.016 | 0,901 ± 0.015 | 0.984 |
| Slot 2 | 0,892 ± 0.017 | 0,884 ± 0.014 | 1.009 |
| Slot 3 | 0,893 ± 0.016 | 0,890 ± 0.014 | 1.002 |
| Slot 4 | 0,892 ± 0.016 | 0,888 ± 0.019 | 1.005 |
| Slot 5 | 0,894 ± 0.016 | 0,886 ± 0.017 | 1.009 |
| Slot 6 | 0,894 ± 0.014 | 0,891 ± 0.017 | 1.003 |
| Slot 7 | 0,887 ± 0.016 | 0,895 ± 0.017 | 0.991 |
| Slot 8 | 0,886 ± 0.016 | 0,897 ± 0.017 | 0.913 |
| Slot 9 | 0,884 ± 0.015 | 0,901 ± 0.018 | 0.981 |
| Slot 10 | 0,883 ± 0.014 | 0,901 ± 0.015 | 0.980 |

Table 11 (above): Normalized estimated complexity, Alice station, 10 slots.

| BOB station | *Kc, L= 24m* | *Kc, L= 1.5m* | $Kc_{(L=24m)}/Kc_{(L=1.5m)}$ |
|---|---|---|---|
| Slot 1 | 0,890 ± 0.014 | 0,903 ± 0.014 | 0.986 |
| Slot 2 | 0,896 ± 0.017 | 0,887 ± 0.015 | 1.010 |
| Slot 3 | 0,897 ± 0.016 | 0,893 ± 0.016 | 1.006 |
| Slot 4 | 0,896 ± 0.015 | 0,891 ± 0.018 | 1.006 |
| Slot 5 | 0,898 ± 0.016 | 0,888 ± 0.017 | 1.011 |
| Slot 6 | 0,897 ± 0.015 | 0,893 ± 0.017 | 1.004 |
| Slot 7 | 0,890 ± 0.014 | 0,897 ± 0.019 | 0.992 |
| Slot 8 | 0,888 ± 0.015 | 0,896 ± 0.015 | 0.991 |
| Slot 9 | 0,887 ± 0.016 | 0,902 ± 0.017 | 0.983 |
| Slot 10 | 0,888 ± 0.015 | 0,902 ± 0.015 | 0.984 |

Table 12 (above): Normalized estimated complexity, Bob station, 10 slots.

| ALICE station | *Hm, L= 24m* | *Hm, L= 1.5m* | $Hm_{(L=24m)}/Hm_{(L=1.5m)}$ |
|---|---|---|---|
| Slot 1 | 0,767 ± 0.060 | 0,819 ± 0.030 | 0.937 |
| Slot 2 | 0,758 ± 0.067 | 0,826 ± 0.035 | 0.918 |
| Slot 3 | 0,757 ± 0.068 | 0,824 ± 0.033 | 0.919 |
| Slot 4 | 0,759 ± 0.066 | 0,810 ± 0.033 | 0.937 |
| Slot 5 | 0,764 ± 0.066 | 0,810 ± 0.032 | 0.943 |
| Slot 6 | 0,768 ± 0.056 | 0,809 ± 0.035 | 0.945 |
| Slot 7 | 0,768 ± 0.059 | 0,813 ± 0.033 | 0.945 |
| Slot 8 | 0,770 ± 0.056 | 0,805 ± 0.033 | 0.957 |
| Slot 9 | 0,772 ± 0.057 | 0,816 ± 0.031 | 0.946 |
| Slot 10 | 0,769 ± 0.059 | 0,803 ± 0.031 | 0.958 |

Table 13 (above): Minimum Entropy, Alice station, 10 slots.

| **BOB station** | ***Hm, L= 24m*** | ***Hm, L= 1.5m*** | $Hm_{(L=24m)}/Hm_{(L=1.5m)}$ |
|---|---|---|---|
| Slot 1 | 0,800 ± 0.069 | 0,852 ± 0.042 | 0.938 |
| Slot 2 | 0,789 ± 0.068 | 0,858 ± 0.045 | 0.920 |
| Slot 3 | 0,785 ± 0.068 | 0,855 ± 0.044 | 0.918 |
| Slot 4 | 0,791 ± 0.071 | 0,848 ± 0.048 | 0.933 |
| Slot 5 | 0,794 ± 0.069 | 0,849 ± 0.047 | 0.935 |
| Slot 6 | 0,802 ± 0.063 | 0,849 ± 0.045 | 0.945 |
| Slot 7 | 0,802 ± 0.063 | 0,847 ± 0.047 | 0.947 |
| Slot 8 | 0,802 ± 0.062 | 0,840 ± 0.050 | 0.955 |
| Slot 9 | 0,804 ± 0.062 | 0,851 ± 0.044 | 0.945 |
| Slot 10 | 0,805 ± 0.068 | 0,842 ± 0.049 | 0.956 |

Table 14 (above): Minimum Entropy, Bob station, 10 slots.